# Stochasticity and traffic jams in the transcription of ribosomal RNA: Intriguing role of termination and antitermination

Stefan Klumpp[*] and Terence Hwa

*Center for Theoretical Biological Physics and Department of Physics,*

*University of California at San Diego, La Jolla, CA  92093-0374*

[*]To whom correspondence should be addressed. *e-mail: klumpp@ctbp.ucsd.edu*

**Corresponding Author:**
Dr. Stefan Klumpp,
Center for Theoretical Biological Physics,
University of California at San Diego,
9500 Gilman Drive, La Jolla, CA 92093-0374
phone: 858 534-7256; fax: 858-534-7697;
email: klumpp@ctbp.ucsd.edu

# ABSTRACT

In fast growing bacteria, ribosomal RNA (rRNA) is required to be transcribed at very high rates to sustain the high cellular demand on ribosome synthesis. This results in dense traffic of RNA polymerases (RNAP). We developed a stochastic model, integrating results of single-molecule and quantitative *in vivo* studies of *E. coli*, to evaluate the quantitative effect of pausing, termination, and antitermination on rRNA transcription. Our calculations reveal that in dense RNAP traffic, spontaneous pausing of RNAP can lead to severe "traffic jams", as manifested in the broad distribution of inter-RNAP distances and can be a major factor limiting transcription and hence growth. Our results suggest the suppression of these pauses by the ribosomal antitermination complex to be essential at fast growth. Moreover, unsuppressed pausing by even a few non-antiterminated RNAPs can already reduce transcription drastically under dense traffic. However, the termination factor Rho can remove the non-antiterminated RNAPs and restore fast transcription. The results thus suggest an intriguing role by Rho to *enhance* rather than attenuate rRNA transcription.

# INTRODUCTION

The *rrn* operons encoding ribosomal RNA (rRNA) in fast growing bacteria are among the most highly transcribed genes known. Fast growth requires a high rate of protein synthesis, which is achieved by a high ribosome concentration in the cell (1). *E. coli* cells which grow with a doubling time of 20 minutes, for example, have on average ~73000 ribosomes per cell (2, 3). To maintain this high ribosome concentration against fast dilution due to cell growth, rRNA has to be synthesized at very high rates, which are achieved by a combination of large copy number of the *rrn* operons and a high transcription rate per operon. The latter is estimated to be 68 transcripts per minute at a growth rate of 20 minutes per doubling (3). For comparison, genes encoding mRNA are typically transcribed at rates of 1-10 per minute (4, 5).

As a consequence of their very high transcription rates, the *rrn* operons exhibit a very high density of transcribing RNA polymerases (RNAPs) as visualized by electron microscopy (6). Despite possible "traffic jams" at high densities, RNAPs transcribe rRNA with high elongation speed of 80-90 nt/sec (7-10), substantially higher than the elongation speed of 35-55 nt/sec known for typical mRNAs (9-11). The higher elongation speed for rRNA transcription was shown to depend on the association of the RNAP with a special antitermination (AT) complex (10), which consists of the Nus proteins, several ribosomal proteins, and a loading sequence (box A) located at the beginning of the rRNA transcript and repeated in the spacer region between the genes encoding the 16S and 23S rRNA (12, 13), see Fig. 1A. The higher elongation speed is attributed to the suppression of RNAP pausing (12), which occur rather frequently, at least *in vitro*, as revealed by both bulk (14) and single molecule experiments (15-18). The physiological implication of the high elongation speed is however not clear.[1] While one may speculate about a connection between the high speed and the high transcription rate, we note that the transcription rate is usually thought to be controlled by the rate of transcript initiation and the speed of elongation is not considered to have an effect on the transcription rate.

---

[1] Faster elongation could contribute to suppress Rho-dependent transcription termination (discussed below), as Rho is less likely to catch up to a faster RNAP (19); but the speed-up effect appears not to be essential for suppressing termination (see Supporting Text for a discussion).

The AT complex suppresses premature termination by the termination factor Rho (12, 20, 21), which binds to the transcript at specific *rut* sites, translocates along it and stops transcription when reaching the RNAP (19, 22). The location of *rut* sites within the *rrn* operons is not known, but studies of various mutants defective in AT (23-26) suggest that that termination occurs predominantly in regions immediately downstream of both box A sequences, see Fig. 1A and Supporting Text. The physiological role of Rho-dependent termination for rRNA transcription is not clear. In the case of mRNA, Rho ensures the coupling of transcription and translation via the 'polarity' effect (27), i.e., by terminating untranslated transcripts. This is however not applicable to rRNA since it is not translated. On the other hand, it seems unlikely that Rho-dependent termination on *rrn* operons is simply a consequence of an unavoidable weak binding of Rho to some generic untranslated transcript, since termination on the *rrn* operons occurs at rather well-defined locations as mentioned above. Furthermore, unlike other termination-antitermination systems, which play important regulatory roles (28), termination and antitermination apparently have no role in regulating the expression of the *rrn* operons (12). Thus, the functional role of Rho-dependent termination (and hence the accompanying need of antitermination) in rRNA transcription is an open question.

In this study, we developed a stochastic model of transcriptional elongation, and applied it to the dense RNAP traffic conditions that dominate the transcription of *rrn* operons at fast growth rates. Our model is based on the quantitative characteristics of transcription *in vivo* and on the dynamics of individual RNAPs as observed *in vitro* (15-18). Individual RNAPs are observed to move in an asynchronous fashion by stochastic single-nucleotide steps, interrupted by different types of pauses. In dense traffic, a trailing RNAP is likely to catch up and push on a paused RNAP. In our model, this pushing has no effect for the majority of pauses (i.e., it does not push forward the paused RNAP), in accordance with observations in single-molecule experiments that the majority of pauses are unaffected by force applied to the RNAP (18, 29). This important feature leads to the possibility of "traffic jams" in dense RNAP traffic since a paused RNAP may force multiple trailing RNAPs to pause as well, hence significantly slowing down the overall rate of transcription.

We used our model to explore systematically the quantitative effect of pausing, termination, and antitermination on the rate of rRNA transcription. Our results suggest that indeed pausing by any individual RNAP during transcription is likely to cause "traffic jams" that would severely restrict the maximal transcription rate attainable under otherwise optimal conditions for fast growth. In particular, we predict that *rRNA could not be transcribed at the high rates necessary to sustain fast growth if the pauses were not suppressed by the AT complex*. This theoretical result suggests that AT plays a crucial role for fast growth even in the absence of Rho-dependent termination and that an *essential* function of the AT complex is *anti-pausing*, which is needed to increase the maximal transcription rate.

However even in the presence of AT, not all RNAPs are antiterminated (30). Therefore we further studied the effect of pausing by a small fraction of non-antiterminated RNAPs amidst the majority of RNAPs with AT. Our results suggest that the effect of pausing on transcription would be so strong in the dense traffic conditions that rare traffic jams caused by the few non-antiterminated RNAPs could already affect the high rate of rRNA synthesis demanded at rapid growth. We then investigated possible effects of Rho-dependent termination on RNAP traffic, and obtained the surprising prediction that higher transcription rates would be attained if Rho could effectively remove the small portion of the stalled, non-antiterminated RNAPs. Our findings suggest an intriguing function for the Rho-dependent termination of rRNA transcription – instead of its commonly understood role in reducing gene expression, Rho may provide a crucial function, under the condition of dense RNAP traffic with a small fraction of non-antiterminated RNAPs, in removing traffic jams and thereby *restoring* the high transcription rates needed to sustain rapid growth.

**RESULTS**

To study transcriptional elongation on highly transcribed genes and, in particular, to investigate the effect of RNAP pausing, we developed a stochastic model of RNAP translocation based on *in* *vivo* and *in vitro* data; see Supporting Text for a detailed description. The model is a variant of stochastic cellular automaton models, which have been studied extensively in statistical physics in the contexts of vehicular traffic (31), non-equilibrium phase transitions (32), and cytoskeletal motors (33, 34). Models of this type have also been used extensively to describe the dynamics of ribosomes on mRNA (35-37) and one recent study used a related model (but without pausing or termination) to study the stepping kinetics of RNAP (38). As depicted in Fig. 1B, the model allows each RNAP (represented by an oval) to be in the active (dark grey) or the paused (white) states. (An extended version of the model, which also incorporates backtracking, the backward translocation of paused RNAPs, is described in the Supporting Text.) Within the active state, each RNAP may step forward by a single base at the rate ε if the next base is not occupied by another RNAP. We refer to ε as the stepping rate or elongation attempt rate; it gives the maximal instantaneous elongation speed $u_0$ of a single RNAP, as $u_0 = \varepsilon\, s$ with the step size $s = 1$ nt. The time-averaged elongation speed $u$ of even a single RNAP is smaller than $u_0$ due to transcriptional pauses. RNAPs are taken to occupy $L$=50 nucleotides on the DNA template (see Supporting Information). If the first 50 nucleotides are not occupied, a new transcript is initiated with rate $\alpha$. We call $\alpha$ the "initiation attempt rate"; the actual initiation rate depends on promoter clearance and is in general smaller than $\alpha$.

In the following, we report results from analytical and numerical studies of the above model. We will describe the behavior of this system step-by-step, first just looking at the effect of RNAP-RNAP interaction, then including the effects of pausing, backtracking, pause suppression by the AT complex, incomplete AT, and finally termination by Rho. For each case, we used model parameters given in Table 1, which are estimated either directly or indirectly from experimental results as described in Supporting Text. We monitored a number of observable transcriptional characteristics, including the average rate of transcription and elongation, and the average density of RNAPs on the transcribed genes, and explored the dependence of these results on key model parameters.

**Dense RNAP traffic without pauses**

We first considered the case of active transcription only, without pausing or termination. When transcription is infrequent, the average rate of transcription $J$, i.e. the number of complete transcripts synthesized per minute by one operon, is just given by the initiation attempt rate $\alpha$ and is independent of the elongation attempt rate $\varepsilon$. As the transcription frequency increases, the RNAP-RNAP interaction (i.e., mutual exclusion) must be taken into account. Naively, one might expect the transcription rate $J$ to be given by $\alpha$ until a critical value $\alpha_c = \varepsilon\, s/L$ set by the clearance rate of the promoter of size $L$, with $J$ taking on the maximal value $J_{max} = \alpha_c$ for $\alpha > \alpha_c$ where the RNAPs are close packed (thin solid lines in Fig. 2A). We refer to the regions at small $\alpha$ as the "initiation-limited regime", and the one at large $\alpha$ as the "elongation-limited regime", since they depend only on the initiation and elongation attempt rate, respectively.

However, due to the stochastic, asynchronous nature of RNAP translocation, neighboring RNAPs affect each other's motion already at densities much below the close-packed limit. This interaction leads to a reduction in the density $\rho$, the average elongation speed $u$, and the transcription rate $J$ compared to the naïve expectation. For the case without pauses, exact analytical results of our model are known (35-37); see Supporting Text. These results are plotted as thick solid lines in Fig. 2 A-C and have been verified by numerical simulations (symbols). Data over a larger range of $\alpha$ are shown in Fig. S1. While the average transcription rate $J$ is indistinguishable from the naïve expectation (thin solid line) at very low initiation attempt rate $\alpha$, it becomes noticeably reduced as $\alpha$ increases. There still exist two distinct regimes of $\alpha$ separated by a critical value $\alpha_c$ (marked by the filled triangle in Fig. S1), with $J$ independent of $\alpha$ in the elongation-dominated regime, $\alpha > \alpha_c$.[2] In this regime the maximal RNAP density is reduced to ~87% of maximal coverage (compare the thick and thin solid lines in Fig. 2B). The average elongation speed, $u(\alpha) \equiv J \cdot L / \rho$, is also diminished for increasing $\alpha$ (Fig. 2C), but only by a small amount (~12% for the minimal speed $u_{min}$) compared to the maximal elongation speed $u_0$.

---

[2] In the context of driven diffusive systems, the two initiation- and elongation-limited regimes are usually referred to as the low-density phase and the maximal-current phase. In addition, the model exhibits a high-density phase, which is not relevant for transcription, as termination of transcription is not limiting the transcription rate.

Simulations of our model also allow us to take 'snapshots' of the spatial distribution of RNAPs on the operons, similar to the images of transcribed genes taken by electron microscopy (6, 7, 25). From these spatial distributions, we determined histograms of the distance between adjacent RNAPs (Fig. S2). In saturated (elongation-limited) traffic (Fig. S2A) the distribution is very narrow with a strong peak at the minimal possible distance $L$ between RNAPs and a small fraction of longer distances. In dilute traffic (Fig. S2B), the distribution is broader and the maximum of the distribution is at a distance larger than $L$.

**The effect of transcriptional pausing**

We next investigated the effect of transcriptional pausing, which is modeled as a stochastic process that occurs at a rate $f$, and persists for a period $\tau$ on average; see Fig. 1B. According to single molecule experiments, there are two main types of pauses, the 'ubiquitous' or 'elemental' short pauses and prolonged backtracking pauses (18, 39). In our model, we assume the duration of the short pauses ('ubiquitous' or 'elemental' pauses) to be independent of the density of RNAP traffic. This assumption is based on the observation that short pauses are unaffected by mechanical force exerted on the RNAP (18, 29), so that a paused RNAP cannot be pushed back into the active state by a trailing RNAP. In contrast, the duration of the less frequent backtracking pauses, which we study in an extended version of our model in Supporting Text, are found to be significantly shortened in dense traffic (Fig. S3), in agreement with *in vitro* and *in vivo* experiments (11). As the inclusion of backtracking hardly affects the results in the dense traffic regime (Fig. S3), we describe below results for the short pauses only. We also neglect the weak sequence-dependence of the short pauses shown by recent single-molecule experiments (17), as it has almost no effect on the results presented here (Fig. S4).

Qualitatively, the results we obtained from simulations including pausing (dashed lines in Fig. 2A-C and Fig. S1) are similar to the case without pausing (thick solid lines). In particular, the two different regimes can still be distinguished in the presence of pauses. In the initiation-limited regime ($\alpha < \alpha_c$), the transcription rate $J$ is hardly affected by the pauses at all (Fig. 2A). The average elongation speed $u$ is gradually reduced due to pausing (Fig. 2C) and the average density $\rho$ is slightly increased (Fig. 2B), reflecting the fact that slower RNAPs remain on the operon for a longer time.

On the other hand, the results for the elongation-limited regime ($\alpha > \alpha_c$) depend strongly on the presence of pauses: Pauses with the parameters given in Table 1 reduced the maximal transcription rate $J_{max}$ from 92 min$^{-1}$ to 43 min$^{-1}$ (plateau of the thick lines in Fig. 2A, the dependence of $J_{max}$ on the pause parameters is shown in Fig. 2D). Similarly, the values $\rho_{max}$ and $u_{min}$ are substantially reduced (Figs. 2B and C). In particular, the reduction of the elongation speed is much stronger in dense traffic than in the dilute traffic (40 and 12 percent, respectively, compared to the case without pauses, see arrows in Fig. S2C). This result can be attributed to traffic jams that build up behind paused RNAPs and amplify the effect of pausing (Fig. S5). These traffic jams are also reflected in the histograms of RNAP-RNAP distances (Fig. S2C and D), where we observe a strong peak at the minimal distance both in saturated and dilute RNAP traffic.

**The effect of ribosomal antitermination**

As mentioned above, ribosomal RNA operons are transcribed with an elongation speed that is increased approximately two-fold compared to protein encoding genes due to the association of the antitermination (AT) complex with the RNAPs transcribing rRNA (10, 40, 41). This increase in speed is generally attributed to the suppression of pauses (see Supporting Text for a discussion). We used our model to see whether the increased elongation speed may be quantitatively explained by *only* a reduction in the pause duration and what effect this change has on the transcription rate. Our study is predicated on experiments carried out to examine transcriptional characteristics with and without AT *in vivo*. For example, in the experiments of ref. (10), the elongation speed $u$ of an *rrn* operon was found to be 79 nt/sec. This value decreased to 52 nt/sec, similar to the elongation speed of mRNA, upon deletion of the AT sequence *boxA*. Similar reductions have been observed for various Nus factor mutations (40, 41).

We determined the transcriptional characteristics in the presence and absence of AT, using the measured elongation speeds as sole constraints; see the results as bold face entries in Table 2. To do this, we estimated the values of the pause frequency $f$ and the elongation attempt rate $\varepsilon$ to be $f$=0.1 sec$^{-1}$ and $\varepsilon$=100 sec$^{-1}$ and took transcription to be in the elongation-limited regime (see Supporting Text). We further assumed that the pause frequency is unchanged with and without AT as suggested by the related N-dependent AT system (42). We first obtained the value of the pause duration $\tau$ with and without AT by demanding the corresponding elongation speeds to match those measured in ref. (10), i.e., $u_{min}$ = 79 nt/sec with AT and $u_{min}$ = 52 nt/sec without AT. The pause duration is found to be $\tau$ = 1.17 sec in the absence of AT, and 0.23 sec in the presence of AT. The result of 1.17 sec is in good agreement with the pause duration of ~ 1 sec observed in single-molecule experiments (15, 17, 18). Also, the predicted reduction in pause duration to 0.23 sec (about 5-fold) agrees well with that of the related N-dependent AT system (about 5-fold) as estimated from the data of ref. (42). We further used our model to determine the expected maximal value of the transcription rate, $J_{max}$, both in the presence and absence of AT. Our simulations predict that AT increases the maximal achievable transcription rate from 38 min$^{-1}$ to 76 min$^{-1}$ (about 2-fold), see Table 2. With this increase, the transcription rate is boosted to the range needed to sustain the fastest growth [68 min$^{-1}$ for 20 min doubling time (3)].

**The effect of partial antitermination**

So far, we have assumed ideal AT assembly and efficiency, i.e., in the presence of the box A sequence each RNAP is antiterminated and exhibits reduced pauses. It has however been shown that even in the nominal presence of AT, a fraction of RNAPs is stopped by Rho (30). Experiments with multiple termination sites in sequence further showed that essentially each RNAP which reads through one termination site also reads through additional termination further downstream (30). This observation suggests that AT is very effective and rather persistent once an RNAP has become antiterminated, but also that not all RNAPs are antiterminated. Partial AT may have a detrimental effect on transcription since RNAPs that are not antiterminated will make longer pauses and may cause jamming even if the majority of RNAPs are antiterminated.

To model the possible consequences of partial AT on *rrn* transcription, we performed simulations for a *mixture* of RNAPs with and without the AT complex. This is implemented by randomly assigning each RNAP a pause duration of either 0.2 sec (with probability $1-p$) or 1 sec (with probability $p$) at the initiation of transcription. Here, $p$ indicates the fraction of non-antiterminated RNAPs, and the pause duration values are chosen according to those in Table 2; they remain unchanged throughout the duration of transcription for each RNAP. In Fig. 3A, we show the transcription rate $J_{max}$ obtained from this model for different fractions $p$ of non-antiterminated RNAPs (filled circles). The transcription rate exhibits a rather steep decrease for small percentages of non-antiterminated RNAPs. E.g., if 10 percent of the RNAPs are not antiterminated, then the maximal transcription rate is predicted to decrease from 81 min$^{-1}$ to 69 min$^{-1}$, a ~30 percent loss of the potential increase in transcription rate due to AT.

**The effect of Rho-dependent termination**

As the decrease of the transcription rate (filled circles) shown in Fig. 3A is caused by the much slower, non-antiterminated RNAPs, we tested the effect of Rho-dependent termination in this context. We performed simulations by incorporating into our model features mimicking forced termination. We only allowed removal of paused non-antiterminated RNAPs.[3] In one implementation (the "uniform termination model"), we assigned a rate $k_t$ at which any paused non-antiterminated RNAP is removed from the system. In an alternative implementation (the "localized termination model"), we allowed termination (with the same rate $k_t$) only to occur within the first 1000 nt of the operon, where RNAPs on average pause for the first time, to mimic a termination site located in the leader region of the transcript (Fig.1A).

The simulation results obtained for the two models are shown as open symbols in Fig. 3A, for a choice of $k_t$ that corresponds to a termination efficiency of 80%, as appropriate for a weak Rho-dependent termination site (30). Similar results are obtained for stronger termination (not shown). These data show that forced termination *increases* the maximal transcription rate in both models if the fraction $p$ of non-antiterminated RNAPs is not too large. For the uniform termination model (open circles in Fig. 3A), we observe an increase of the transcription rate for $p$ below ~40% (where the curves with the solid and open circles cross each other). For higher fractions of non-antiterminated RNAPs (and thus also in the complete absence of antitermination, $p$=100%), the transcription rate is however found to be strongly reduced, as one would usually expect in the presence of strong indiscriminant termination. For the localized termination model (open squares in Fig. 3A), the transcription rate is increased for up to 90% of non-antiterminated RNAPs. Thus, forced termination near the start of transcription initiation can have a substantially more beneficial effect in speeding up transcription. We attribute this effect to the fact that removing non-antiterminated RNAPs soon after initiation not only reduces the number of traffic obstacles, but that these non-antiterminated RNAPs are also quite likely to be replaced by antiterminated RNAPs; whereas forced termination downstream of the transcription start has the adverse effect of reducing the density of the transcribing RNAP. For both models, the transcription rate can be maintained close to the level of perfect antitermination with as much as 10 percent RNAPs without AT. This conclusion remains valid if we consider a more realistic *rrn* operon (Fig. 1A) where the AT complex is reloaded in the spacer region between the 16S and 23S genes with an additional termination site immediately downstream (not shown).

---

[3] The selectivity of rho-dependent termination is higher in our model than in an alternative scenario where selection is *only* based on the elongation speed (23), see Supporting Text for a discussion.

**Transcription of rRNA in wild type and AT mutants**

Finally we extended our model to the structure of *E. coli rrn* operons as shown in Fig. 1A, where the AT complex is assembled in the leader region of the transcript, followed by a termination site, and reassembled in the spacer region between the 16S and 23S genes, again followed by a termination site (see Methods for the modeling details). We used this full version of our model to study the effect of AT mutants. Quan *et al*. have recently characterized viable NusA and NusB mutants (25). They showed that these strains exhibit polarity between the 16S and 23S genes and quantified this observation using electron microscopy of the *rrn* operons. To see whether our model can account for these observations, we simulated transcription of rRNA with the full model. We chose the initiation attempt rate to be $\alpha$ =120 min$^{-1}$ so that the RNAP density for perfect AT matches the measured density for the wild type (see Table S1, rows 1 and 2). For this value of $\alpha$, RNAP traffic is dense, about 73% saturated (data not shown). The transcription rate predicted by our model is 58 min$^{-1}$ (Table S1 row 2). The transcription rate was not directly measured by Quan et al. (25). However, by using the known relation between growth rate and *rrn* transcription for wild type *E. coli* cells (2, 3), we can predict a corresponding doubling time of ~ 24 min which agrees with the one observed (25). We then varied the degree of AT to mimic defective AT due to NusB or NusA mutations and determined the numbers of RNAPs on the 16S and 23S genes (Table S1), as well as histograms of RNAP-RNAP distances (Fig. 3B and C). We obtain quantitative agreement with the experimental values for the number of RNAPs per gene, if we assume 60% of the RNAPs to be not antiterminated in the Nus mutants (Table S1, rows 3-5). We also determined the corresponding transcription rate (of full length transcripts) and estimated the corresponding doubling time to be ~36-38 min for these mutants (Table S1 row 5). This estimate agrees well with the measured doubling time for the NusA mutant (38 min). The NusB mutant exhibits *rrn* transcription very similar to the NusA mutant, but has a substantially larger doubling time of 58 min, possibly indicating an additional defect of this mutant beyond ribosomal AT.

For the wild-type cells and the NusB mutant, Quan *et al.* also determined histograms of the RNAP-RNAP distances measured in electron microscopy images of the *rrn* operons (25). The corresponding histograms obtained from our model are shown in Fig. 3B and C. For perfect AT (Fig. 3B), we find that the histograms are almost the same for the 16S and 23S genes, with a strong peak at distances of ~90 nt and a very small fraction of RNAP-RNAP distances larger than 250 nt, in good agreement with the histograms of ref. (25) for wild type cells (Table S1). For defective AT with only 40% of the RNAPs antiterminated (Fig. 3C), we find that the distribution for the 23S gene is much broader than the one for the 16S gene, and that the distributions for both genes are broadened compared to the case of perfect AT with an increased fraction of large distances (Table S1), in good agreement with the results from ref. (25) for the NusB mutant.

**DISCUSSION**

The rate of transcription is usually thought to be controlled by the rate of transcript initiation, while the speed of transcript elongation has no effect on the transcription rate. While this view is most likely correct for most bacterial genes, we show here that the speed of elongation also plays an important role for highly transcribed genes such as the *rrn* operons in fast growing bacteria, since it affects the rate of promoter clearance when RNAP traffic is congested. We developed a stochastic model to study such situations, based on the known stochastic dynamics of individual RNAPs (15, 18). In our model, the RNAPs move in an asynchronous fashion and the pausing of one RNAP may impede the stepping of a trailing RNAP. Experimental evidence indicates that backtracked pausing RNAPs may be pushed forward by a trailing RNAP (11). Including this aspect in our model, we find that backtracking pauses should be fully suppressed in dense RNAP traffic (see Supporting Text). However, single molecule experiments show that the majority of pauses (which are short pauses without backtracking) are unaffected by force applied to the RNAP (18, 29) and thus unlikely to be affected by the pushing force of a trailing RNAP. Our results indicate that the transcription rate can be significantly affected by the stochastic nature of the elongation process, and particularly by transcriptional pausing which creates traffic jams and slows down the overall RNAP traffic (Fig. 2).

**The role of antitermination in rRNA transcription**

RNAPs transcribing rRNA in bacteria are modified by an antitermination complex that speeds up transcript elongation and suppresses Rho-dependent termination (12, 20, 21, 40).[4] The increase in elongation speed due to AT is attributed to the suppression of transcriptional pausing (12), but the physiological consequence of this anti-pausing activity is unclear. Our analysis suggest that the suppression of pauses is essential at fast growth where RNAP traffic is dense: The *rrn* operons in fast growing bacteria are highly transcribed, with transcription rates of up to ~70 transcripts per minute per operon (2, 3).[5] According to our analysis, it is not possible to obtain transcription rates of this magnitude in the absence of AT (Table 2), given the known dynamics of elongation with pauses (15, 18) and the speed of transcription measured *in vivo* in the absence of AT [52 nt/sec (10)], because pausing RNAPs in dense traffic induce RNAP "traffic jams".[6] With AT, we find that both the physiologically required transcription rate for fast growth [68 min$^{-1}$ at 20 min doubling time (3)] and the observed increase in elongation speed [to 79 nt/sec (10)] can be quantitatively explained by a 5-fold decrease in the pause duration from ~ 1 sec to ~ 0.2 sec (Table 2). Our results suggest that an essential physiological function of ribosomal AT is *anti-pausing*, designed to increase the transcription rate of rRNA to sustain the demand at rapid growth, rather than to suppress Rho-dependent termination. We predict that mutants with defective AT will exhibit growth defects, mainly at very rapid growth, even if Rho-dependent termination is not operative there, e.g., in Rho or rut site mutants, or if Rho activity is suppressed by an antibiotic (45). For cells defective in both AT and Rho activity, Table 2 makes quantitative predictions for the transcription rate of rRNA and the doubling time for growth in rich medium. We note however that such experiments are not easy to do or interpret, because ribosome synthesis is a core component of cell growth, and perturbations likely have additional effects.

---

[4] The AT complex is likely to have additional functions such as adapting the elongation speed for optimal folding of rRNA (42), facilitating rRNA processing (44) and assembly of the ribosome subunits (13), which are outside the scope of this study.

**The role of Rho-dependent termination in rRNA transcription**

The above interpretation of *rrn* antitermination as an anti-pausing mechanism naturally leads to questions on possible function(s) for Rho-dependent termination in rRNA transcription. Rho-dependent termination ensures the coupling of transcription and translation for protein-encoding genes as evidenced by the polarity effect (27), but its role for rRNA genes has not been addressed previously. Our simulations indicate that the rate of rRNA transcription can be significantly reduced by the presence of even a small fraction (e.g., 10%) of non-antiterminated RNAPs, which make longer pauses (Fig. 3). Since even in the presence of AT, not all RNAPs are antiterminated (30), such non-antiterminated RNAPs are expected to appear in the transcription of rRNA. The high transcription rates needed for fast growth thus require a mechanism to remove these obstacles. We propose that Rho-dependent termination can provide such a mechanism (Fig. 3): Assuming that Rho can efficiently terminate those RNAPs that escaped antitermination, e.g., removing a large fraction of them during their longer pauses, results of our simulations predict that the reduction in transcription can be almost entirely restored, unless the fraction of non-antiterminated RNAPs is so large that removing them lead to significant reduction in RNAP density. We thus conclude that Rho-dependent termination could actually *increase* rather than decrease transcription as one would naively expect. In this way, Rho-dependent termination may be recruited to work as an integral part of a transcriptional system designated to achieve high transcription rates for rRNA.

---

[5] Even higher transcription rates are necessary in strains with reduced numbers of *rrn* operons, see Supporting Text for a discussion.
[6] The transcription rates without AT are however sufficient to account for the transcription of highly transcribed protein-encoding genes such as those encoding ribosomal proteins (5).

**ACKNOWLEDGEMENTS**

SK was supported in part by a fellowship from Deutsche Forschungsgemeinschaft (Grants KL818/1-1 and 1-2). SK and TH are grateful to further support by the NSF through Grant PHY-0822283 to the PFC-sponsored Center for Theoretical Biological Physics, and through Grant MCB0746581 to TH.

**FIGURE LEGENDS**

**Figure 1: Modeling transcription of rRNA:** (A) Schematic structure of an *rrn* operon with the genes encoding the 16S, 23S, and 5S rRNAs (tRNA genes that are present in some *rrn* operons are not shown) showing the loading sequence boxA of the antitermination complex (black ovals), the approximate positions of rho-dependent terminators as inferred from studies of mutants defective in antitermination (T; see Supporting Information), and the promoter pair P1-P2 (P). (B): Model for the traffic of RNA polymerases. Active RNAPs (dark grey) elongate an RNA transcript by making single-nucleotide forward steps along the DNA template. These elongation steps occur with rate $\varepsilon$, provided the site in front of the RNAP is not occupied by another RNAP. In addition, RNAPs can switch to an inactive or paused state (white). The transition rate to the paused state is given by the pause frequency $f$ and a paused RNAP returns to the active state with rate $1/\tau$, where $\tau$ is the average pause duration. Paused RNAPs are terminated with rate $k_t$ mimicking Rho-dependent termination. At the promoter (represented by the first site to the left) transcripts are initiated with the initiation attempt rate $\alpha$ if the access of the initiating RNAP to the promoter is not blocked by another RNAP bound there.

**Figure 2: Effect of pauses during transcription.** (A) Transcription rate $J$, (B) RNAP density $\rho$, and (C) elongation velocity $u$ as functions of the initiation attempt rate $\alpha$. Thin solid lines indicate the naïve (and incorrect) estimate in which RNAPs move synchronously. Thick solid lines (analytical results as given by eqs. [1]-[3] in Supporting Text) and squares (simulations) show results for the case without transcriptional pausing. Circles show the case including pauses with pause parameters from Table 1. (D) The maximal transcription rate $J_{max}$, which is attained for large initiation attempt rates $\alpha$, as a function of the pause parameters $\tau$ and $f$. The arrow indicates the parameters used in A-C.

**Figure 3: Effect of partial antitermination.** A: The maximal transcription rate $J_{max}$ as obtained from simulations with two populations of RNAPs with different pause durations $\tau= 0.2$ sec and $\tau=1$ sec, which represent antiterminated and non-antiterminated RNAPs, respectively (filled symbols). The effect of AT is reduced by the presence of non-antiterminated RNAPs. Termination of the non-antiterminated RNAPs (with a termination efficiency of 80 % during pauses, open symbols) can compensate this decrease for small fractions of non-antiterminated RNAPs. B and C: Histograms of RNAP-RNAP distances for perfect (B) and defective AT (C). Distances between adjacent RNAPs were determined from snapshots of the spatial distribution of RNAPs on simulated *rrn* operons as shown in Fig. 1A. Black and white bars show distributions for the 16S and 23S genes, respectively. For perfect AT, the distributions are almost the same, for defective AT the distribution for the 23S gene is much broader than that for the 16S gene, similar to what has been observed for a NusB mutant (25, Fig. 3).

**Table 1: Model parameters and estimated values.**

|                          | Symbol | Value [*]            |
|--------------------------|--------|----------------------|
| Elongation attempt rate  | $\varepsilon$ | 100 sec$^{-1}$ |
| Pause frequency          | $f$    | 0.1 sec$^{-1}$       |
| Pause duration           | $\tau$ | 1 sec                |
| Size of RNAP footprint   | $L$    | 50 nt                |
| Termination rate         | $k_t$  | 2 – 10 sec$^{-1}$    |

(*) See Supporting Information for a detailed description of how these values were estimated.

**Table 2: Quantitative effect of antitermination as predicted by the model.** The elongation speeds are taken from ref. (10). Bold-face numbers indicate the predicted transcriptional characteristics at the (predicted) elongation attempt rate of $\varepsilon = 100$/sec; see the text for details.

|  | without AT | with AT | fold change |
| --- | --- | --- | --- |
| Elongation speed $u$ | 52 nt/sec | 79 nt/sec | 1.5x |
| Pause duration $\tau$ | **1.17 sec** | **0.23 sec** | **5x** |
| Transcription rate $J_{max}$ | **38 min$^{-1}$** | **76 min$^{-1}$** | **2x** |

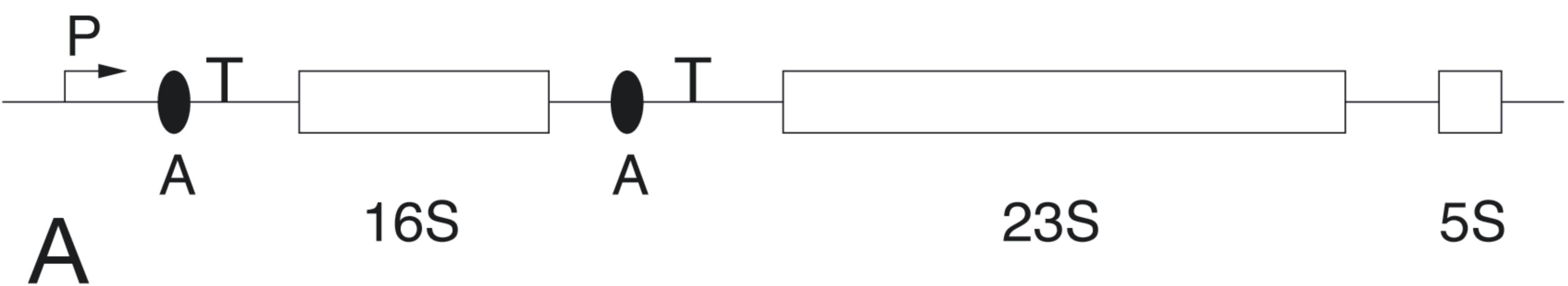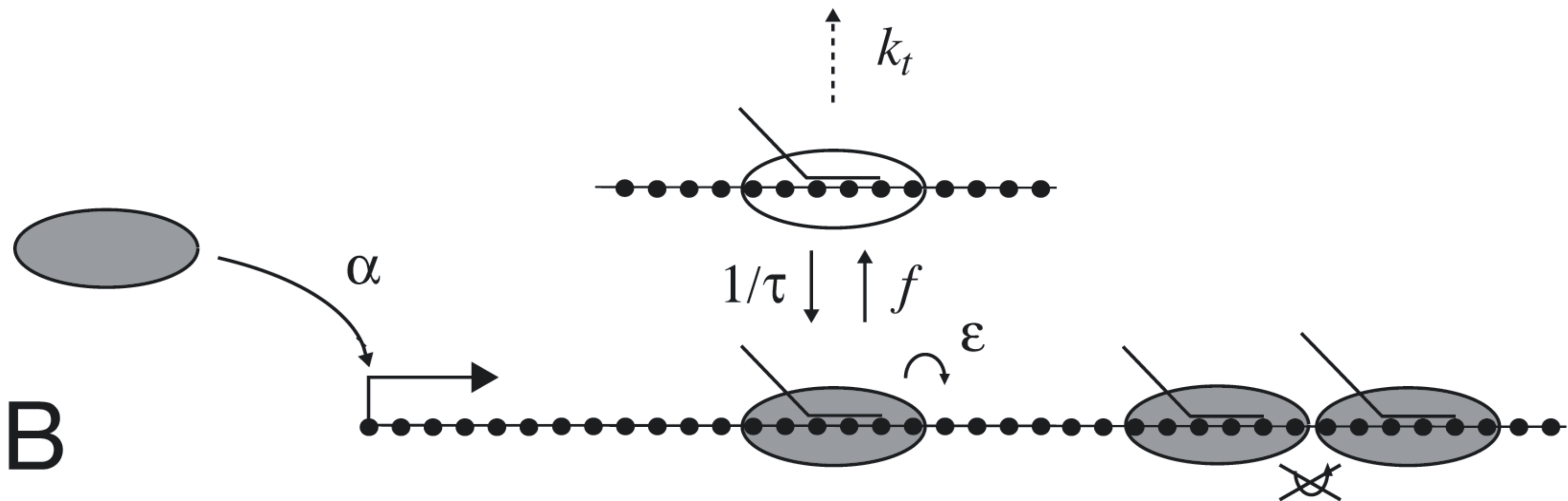

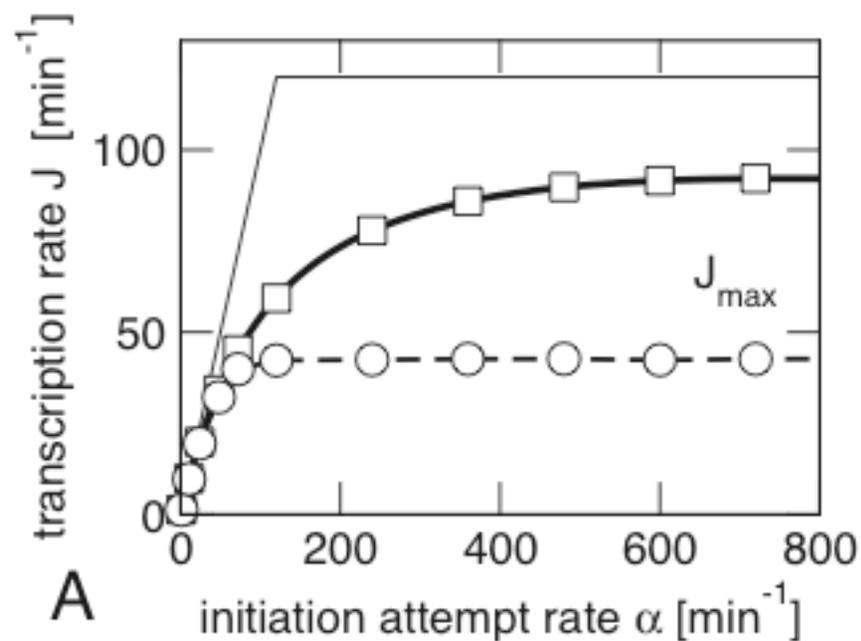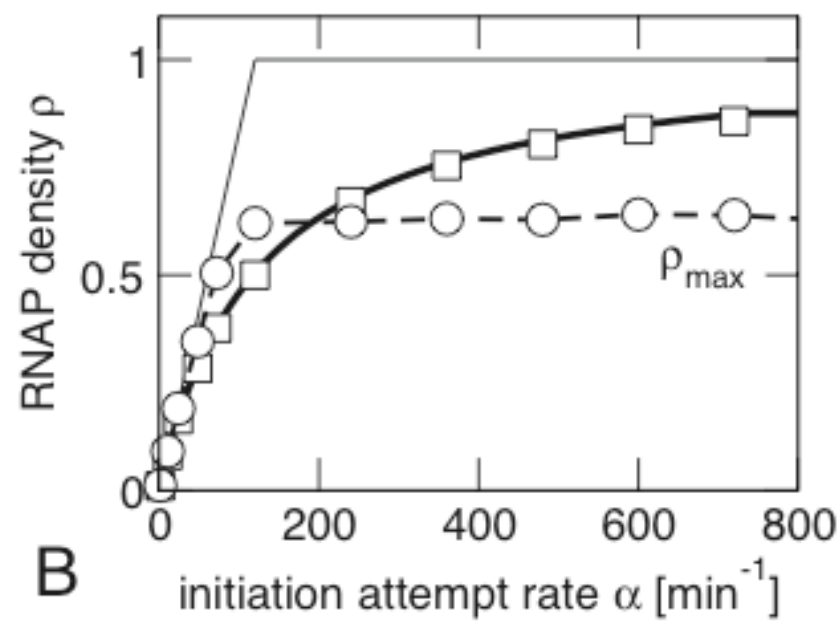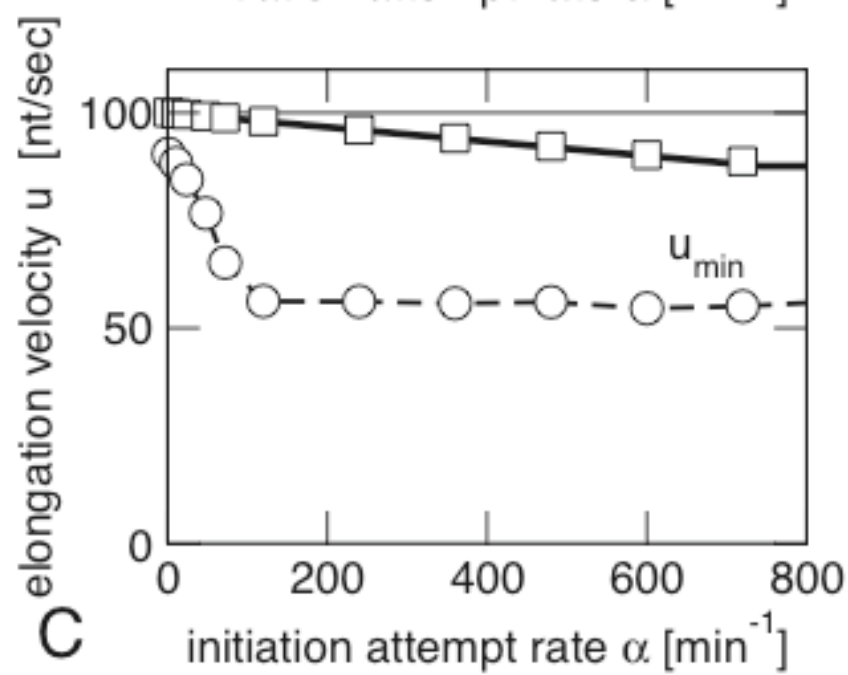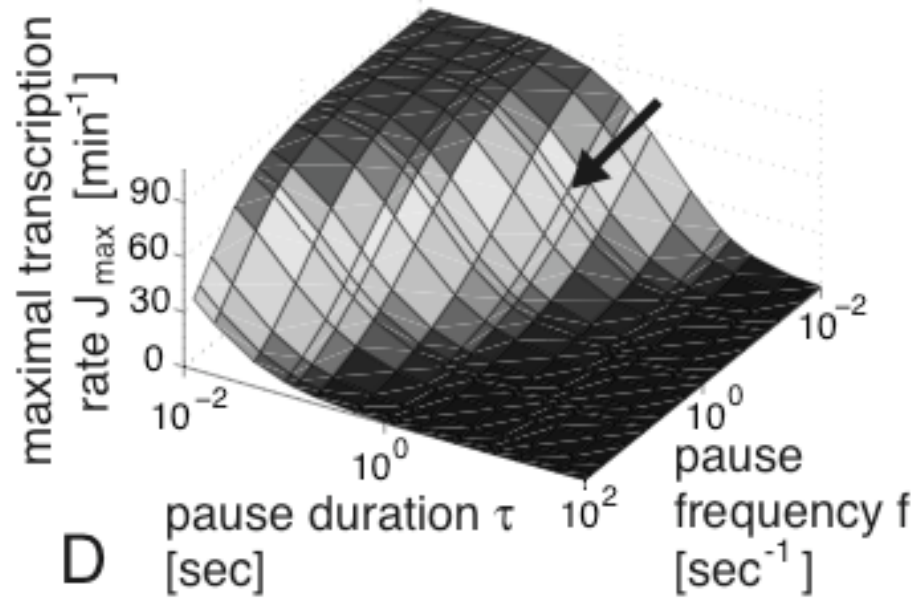

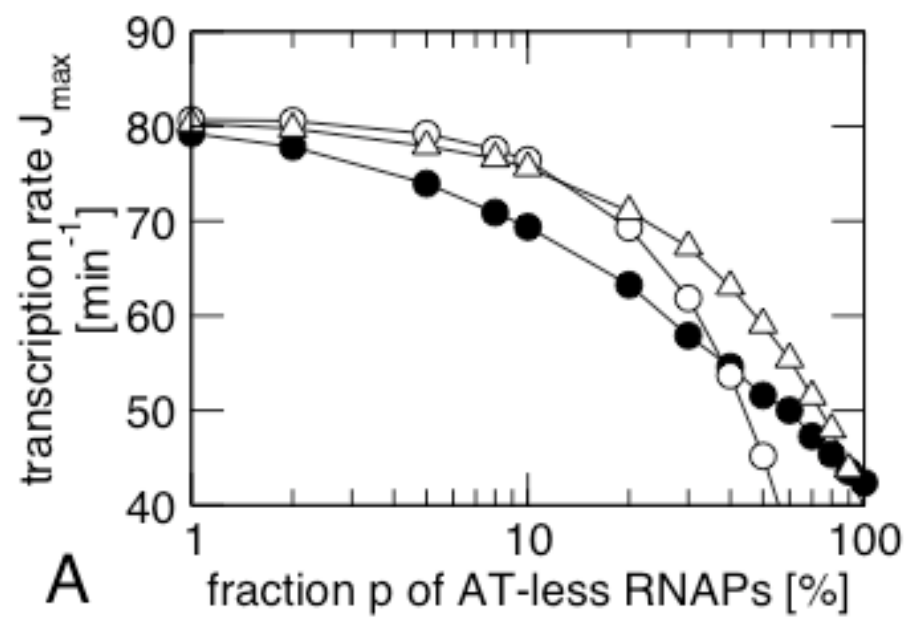
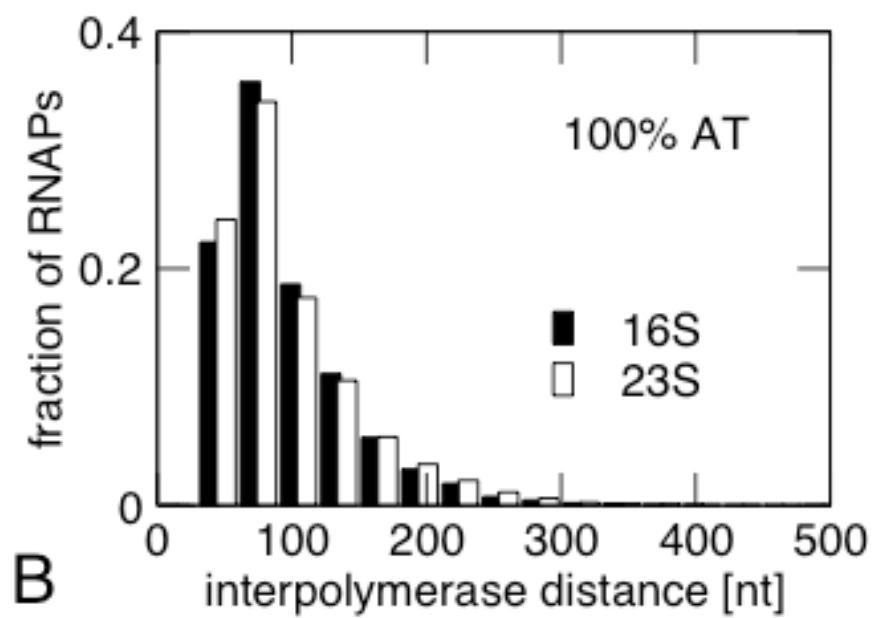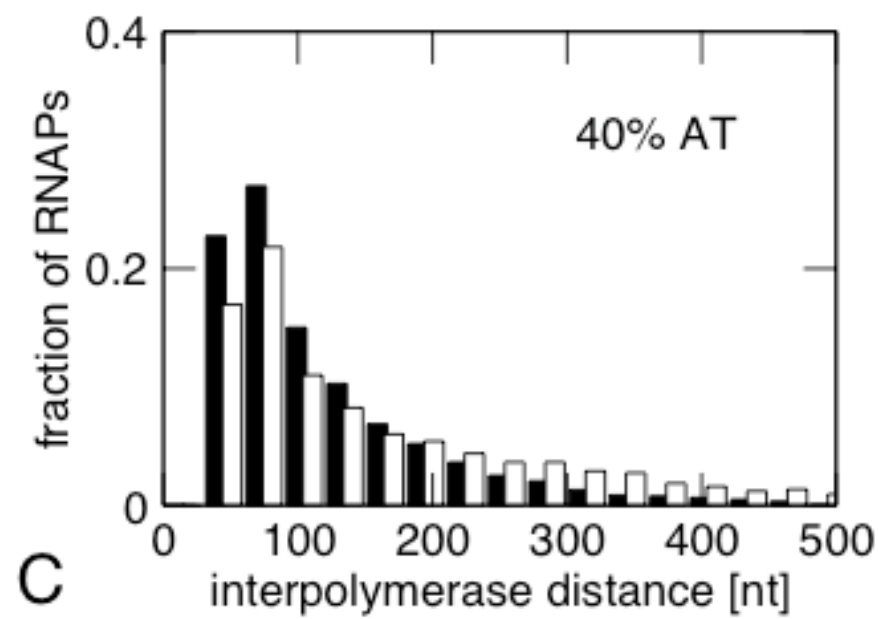

# Stochasticity and traffic jams in the transcription of ribosomal RNA: Intriguing role of termination and antitermination

Stefan Klumpp and Terence Hwa

SUPPORTING TEXT

A. Methods

**The model**

The transcription by RNAPs in dense traffic is modeled by a stochastic cellular automaton model (1), in which the RNAPs are represented as extended objects of size $L$ confined to move along a one-dimensional lattice; sites of this lattice represent the individual bases of a DNA template. Each RNAP can be in either the active, paused, or backtracked state independently of the state of other RNAPs. An active RNAP transcribes RNA by making a single nucleotide step forward at a stepping or elongation attempt rate $\varepsilon$, provided that the access to the next base is not blocked by the presence of another RNAP. During transcription, the active RNAP may switch stochastically to the paused state with a rate $f$ (see section B for a discussion of the stochasticity and sequence-dependence of pauses). A paused RNAP remains at the same site. It can switch back to the active state with rate $1/\tau$, so that the average duration of a pause is $\tau$. In an extended version of our model, we also included backtracking of paused RNAPs, which has however little effect on transcription in dense traffic (see section C below). Finally, for simulations of rho-dependent termination, termination is implemented by removing non-antiterminated, paused RNAPs with rate $k_t$. This implementation mimics the effect that rho can only displace the non-antiterminated RNAPs when it catches up with those RNAPs in the paused state (2). Localized termination sites were implemented by allowing this termination process only within a template stretch of 1000 nt, so that RNAPs pause there on average once. In simulations with partial AT, each initiating

RNAP was randomly chosen to be either antiterminated (with probability *p*) or not antiterminated (with probability 1-*p*). In the simulations that included reloading of the AT complex in the spacer region, this assigment was renewed with the same probabilities at site 2000. In this case, termination was allowed at the first 1000 sites and between sites 2000 and 3000 to mimic the two termination sites following the leader and spacer boxA sequences, respectively (indicated by T in Fig. 1A and discussed in section E below).

The initiation of transcription is described by the initiation attempt rate α with which RNAPs are inserted at the first *L*=50 sites provided that these sites are not occupied. This rate is taken to summarize all processes of the initiation stage such as RNAP binding to the promoter, open complex formation, and initiation of transcript elongation (3). Actively transcribing RNAPs that reach the end of the operon are terminated, i.e. removed, when making a forward step from the last site. Fast termination is implied by the absence of RNAP jamming at the end of the operon in a recent electron microscopy study of ribosomal RNA operons (4). We estimated the values for all model parameters from *in vitro* and *in vivo* data, see section B below. The estimated values are summarized in Table 1.

Our model is a variant of stochastic cellular automata or driven diffusive systems and incorporates several features that have been extensively studied in the non-equilibrium statistical physics literature. In particular, models with extended particles, i.e. particle that occupy more than a single lattice site, have been studied in detail, mostly motivated by the traffic of ribosomes on mRNA (5-7), and we make use of some known analytical results for the case without pausing (see below). Pauses of the type observed for RNAPs has not been studied within these models, but some well-studied models exhibit quite similar effects. In particular, sequence-dependent stepping rates have been investigated extensively and also lead to the build-up of traffic jams behind a slow site (8). We note however that in such a model all particles move slowly at a site with a low stepping rate, while in our model only a fraction of RNAPs pauses at each pause site (see below for a description our simulations of sequence-dependent pauses). Premature termination as included in our model is related to unbinding of particles in models used to describe the traffic of cytoskeletal motors (9, 10); however, in these models unbinding typically occurs together with binding of particles to the track, which does not apply to RNAPs that initiate transcription from the promoter rather than from random positions along the DNA template.

**Model simulations and analytical results**

The model is simulated using discrete Monte Carlo steps that correspond to $5\times10^{-3}$ s. Each step consists of $N$ moves where $N$ is given by the length of the lattice or the number of nucleotides in the operon, for which we used the length of the *rrnC* operon, $N\approx5450$ (4). At each of these moves, a site is chosen randomly and updated according to the described rates (11). All steps that would move an RNAP to a site occupied by another RNAP are rejected. Simulations were run for $5\times10^6$ steps and data for steady-state average were taken starting after $5\times10^5$ steps.

In the limit where pauses are absent, our model reduces to the well-studied asymmetric simple exclusion process with extended particles for which a number of analytical results are known (5-7), shown as thick solid lines in Fig. 2 A-C and Fig. S1. The transcription rate corresponds to the current in that model and is given by

$$J(\alpha)=\alpha(\varepsilon-\alpha)/[\varepsilon+\alpha(L-1)] \quad \text{and} \quad J_{max}=\varepsilon/(1+L^{1/2})^2 \qquad [1]$$

in the initiation limited and in the elongation limited situation, respectively. These two regimes are separated by a continuous phase transition that occurs for $\alpha_c=\varepsilon/(1+L^{1/2})$. The RNAP density is given by

$$\rho(\alpha)=L\alpha/[\varepsilon+\alpha(L-1)] \quad \text{and} \quad \rho_{max}=1/[1+L^{-1/2}], \qquad [2]$$

respectively. From these two known results, we obtain the elongation velocity $u$ using the relation $u=JL/\rho$, which leads to

$$u(\alpha)=\varepsilon s\,(1-\alpha/\varepsilon) \quad \text{and} \quad u_{min}=\varepsilon s/(1+L^{-1/2}) \qquad [3]$$

in the initiation-limited regime and at maximal transcription, respectively.

**B. Parameter estimation**

In this section, we describe in detail how we estimated values for the parameters of our model using *in vivo* and *in vitro* data from the literature. The parameter values are summarized in Table 1.

**Pauses:**

The frequency $f$ and duration $\tau$ of pauses have been directly measured *in vitro* in single molecule experiments (12-16) Experiments with high resolution show that RNAPs exhibit both short and prolonged pauses (discussed below). The majority of pauses are short (12, 13). These pauses occur with a frequency of 0.07-0.15 $sec^{-1}$ and have durations of approximately 1 sec (12-14). *In vivo* values for these parameters have not been measured. However, we expect these *in vitro* values also to be representative for the situation in vivo, despite the fact that transcription *in vivo* proceeds faster than under the conditions of the single molecule experiments. The rationale for this assumption is given by the following two arguments: (i) The pause frequency has been observed to be constant over the range of elongation rates accessible *in vitro* (14), so it may be extrapolated to the higher elongation rate *in vivo*.[1] (ii) Assuming a stepping rate of 100 $sec^{-1}$ (see below), the *in vitro* pause parameters are consistent with the *in vivo* elongation speeds of mRNA (17-19) and rRNA in the absence of antitermination (18). In fact, when we determined the pause duration from the elongation speeed of rRNA without antitermination, we obtained values for the pause duration $\tau$ between 1 and 2 seconds within the expected range of stepping rates, see Table 2 and section D below.

In single-molecule experiments, pauses occur in a stochastic fashion. The frequency of pauses exhibits rather small variation along the DNA template when measured with ~100 nt resolution (13), and both the distance and time between subsequent pauses follow single-exponential distributions (12). Recent experiments with single-base resolution revealed a weak sequence dependence of the short pauses (14): Pauses occur preferentially at specific sites on the template, but only a small fraction of the RNAPs pauses at any given pause site. We have checked that pauses at discrete sites (randomized by the fact that not all RNAPs pause at every pause site) with an average density of one pause site per 100 nt (14) lead to almost the same results as completely random pausing at all sites (Fig. S4). Results for the maximal transcription rate and the corresponding elongation speed obtained for different sequences with randomly chosen pause sites are distributed around the value for random pausing and deviate from it by at most ~15 and ~20 percent, respectively (Fig. S4B and D). In the studies described in the main text, we thus neglected the sequence-dependence of the short pauses in order to keep the model simple. Prolonged backtracking pauses, on the other hand, exhibit a strong sequence-dependence, which is taken into account in our simulations of the extended model, where backtracking was incorporated explicitly (see below).

---

[1] We note that there is kinetic competition between pausing and translocation. Since the elongation rate *in vivo* is higher than what is measured *in vitro*, the probability of pausing is reduced compared to what is measured *in vitro*. The same pause frequency (transition to pauses per time interval) thus corresponds to a smaller pause density (pauses per length transcribed).

**Stepping rate:**

The stepping rate has been measured *in vitro*, where it ranges between 5 and 30 nt/sec (12-14) and is not known *in vivo*. The *in vivo* value however has to be larger than the values measured *in vitro*, because typical elongation speeds measured *in vivo* are in the range of 20-80 nt/sec (17-20). These speeds are average speeds, slowed down due to pauses, so that the stepping rate $\varepsilon$, which corresponds to the elongation speed in the absence of pauses, should be even larger. Lower bounds for the stepping rate can thus be obtained from the elongation speeds of transcripts, on which pausing is suppressed by an antitermination system. One such system is rRNA, which is transcribed at 80-90 nt/sec in dense RNAP traffic (17, 18, 20, 21), from which we obtain a lower bound on the stepping rate $\varepsilon$ of 90-100 sec$^{-1}$. Another system is the late transcript of phage $\lambda$, which is transcribed by RNAPs antiterminated by the Q protein. In this case, the elongation speed has been measured to be 90-100 nt/sec (22), however in this case in a dilute RNAP traffic situation. This also leads to an estimate for the lower bound of $\varepsilon$ of 90-100 sec$^{-1}$. [An even higher elongation speed of 135 nt/sec has been observed in one experiment (21) for an E. coli strain with reduced number of *rrn* operons. The mechanism that leads to this unusually high elongation speed is unknown, see also the discussion in section F below.] The agreement of the estimates obtained from different mechanisms of pause suppression systems provides a first hint that the observed elongation speeds are close to the maximally possible values and that pausing may be almost completely suppressed in these systems. Our analysis of the effect of ribosomal AT also implies an upper bound on the stepping rate (described in section

D below) at ~110-150 sec$^{-1}$. We therefore expect the stepping rate *in vivo* to be ~100 sec$^{-1}$. The latter value has been used in all simulations described in the main text, but different values in the estimate range lead to similar predictions for the effect of ribosomal AT (see section D below and Fig. S6).

**Size of the RNAP footprint:**

The number of nucleotides $L$ occupied by a single RNAP is estimated from DNase footprinting experiments (23-27). These experiments exhibit footprints of 77 and 50 nt for promoter-bound RNAPs in open and closed complexes, respectively. These footprints reach from positions -57 to +20 for the open complex (23, 24) and from -55 to -5 for the closed complex (25, 26). An elongating RNAP has a footprint of 32 nt (23). The footprint of an elongating RNAP that is stalled at position +32 (i.e. after synthesizing a transcript of 32 nt) reaches to +45; in this situation a second RNAP can bind to the promoter (27). This observation provides a lower limit for the distance an RNAP must have moved to allow the initiation of the next transcript. However, in this situation, the trailing RNAP may not yet be able to form an open complex, since the footprints of the elongating leading RNAP (+13 to +45) and the trailing promoter-bound open complex RNAP (-57 to +20) would overlap by ~7 nt. This means that the distance an RNAP has to move away from the promoter before a new transcript can be initiated is probably larger than ~40 nt. An even larger value of 50 nt has been estimated from the occlusion of the ribosomal RNA promoter P2 due to RNAPs initiating transcription from the upstream promoter P1 (28), which has been determined by comparing the transcription rates from the individual promoters P1 and P2 and the combined promoter P1-P2. In addition, RNAP is much larger than DNase I, so that positions that are accessible to DNase I may not be accessible to another RNAP. This implies that DNase footprints may possibly underestimate the volume from which a trailing RNAP is excluded by the presence of the leading RNAP. For elongating RNAPs another reason why DNase footprint may underestimate the *in vivo* size of the elongation complex is the presence of elongation or antitermination factors, which may increase the size of the complex compared to the values measured in vitro in the absence of these factors. In our simulations we thus used a footprint size of $L$=50, slightly larger than the values measured *in vitro*, but in agreement in the *in vivo* estimate of ref. (28). Using the smaller size $L$=35 nt increases the maximal transcription rate in the absence of pauses (compare eq. [1] above), but has a smaller effect in the presence of pauses.

**Termination:**

In the simulations that included termination, a paused RNAP may be removed with the termination rate $k_t$. This mimics the fact that the termination factor rho, which binds to the transcript at specific sites (*rut* sites) and translocates along it in a directed fashion (2), may catch up with the elongating RNAP when the latter pauses as indicated by the observation that the sites of termination correlate very well with pause sites (29, 30). We determined the termination rate $k_t$ from the probability of termination or the termination efficiency $T$ of rho-dependent termination sites. The termination efficiency is given by the relative frequency of termination compared to escape from the paused state, i.e. by $T=k_t/(k_t+1/\tau)$. Rho-dependent termination sites have termination efficiencies between 0.7 and 0.9 (31), which leads to values of $k_t$ between 2.3 s$^{-1}$ and 19 s$^{-1}$. In most cases, we assumed termination to have intermediate strength and took $T$ to be 0.8, which corresponds to $k_t= 4$ s$^{-1}$.

In our model, antitermination has two distinct effects: suppression of pauses and protection against rho-dependent termination. In general, these two effects are intimately linked and reduced pausing should contribute to the suppression of termination through "kinetic competition", as it is less likely for Rho to catch up to the faster RNAPs (32). These two effects have however been disentangled in experiments with the two AT systems of phage λ, the N-dependent AT system which is very closely related to the *rrn* AT systems, and the Q-dependent AT system (33-35). In both cases, the AT complex has been shown to suppress pausing (36, 37), but also to provide protection against Rho beyond simple kinetic competition: In vitro experiments show that these AT complexes also protect RNAPs that are slowed down artificially by low nucleotide concentrations (33-35), which suggest that AT also stabilizes RNAPs against the action of Rho once it has reached the RNAP and that the suppression of pausing is not essential for the suppression of Rho-dependent termination.[2] In our model, we therefore do not allow the termination of paused antiterminated RNAPs in our model. We note that in the situation of dense RNAP traffic with a mixture of slow non-antiterminated and fast antiterminated RNAPs, the additional AT-derived protection against Rho (beyond suppressed pausing) increases the selectivity of Rho to exclusively remove the slow non-antiterminated RNAPs compared to pure kinetic competition, i.e. selection based only on the duration of pauses..

---

[2] In the case of Q-dependent AT from the λ-related phage 82, it has also been shown that the suppression of termination once Rho reaches the RNAP requires the NusA protein, but the suppression of pauses does not (35).

**Backtracking:**

The extended version of our model, in which we also included backtracking (Fig. S3) has a number of additional parameters, which characterize the frequency and duration of prolonged pauses and the dynamics of backtracked RNAPs. Single molecule experiments indicate that backtracking pauses are relatively rare and represent a small fraction (< 5 or 10 percent) of all pauses (12, 13). They are distinguished from pauses without backtracking by pulling on the RNAP. While the duration of the short pauses is unaffected by force, the duration of backtracking pauses increases if an opposing force is applied and decreases if an assisting force is applied (13, 16, 38). Histograms of the observed pause duration have been fitted by double exponentials (12, 13). Whether the two time scales, which characterize these distributions, represent the durations of short pauses and backtracking pauses or two types of short pauses is not completely clear, and different interpretations have been advanced by different groups (12, 14). In the first case (12, 39), pauses longer than ~10 sec would be considered as backtracking pauses. In the latter case, only those pauses that are not described by the double exponential distribution would be considered as backtracking pauses (13, 14). These pauses have durations of > 25 sec (13, 14). In both cases, however, backtracking pauses would represent less than 5-10 percent of all pauses. In addition, the densities of pause sites with long pause durations on the template reported by several groups are quite similar, typically one prolonged pause is observed every 1000-3000 nt (12, 13, 16). In contrast to short pauses without backtracking, backtracking pauses are strongly sequence dependent (12, 14-16). This implies that the precise number of backtracking pause sites depends on the template that is used. For a template of length ~5000 nt, we expect to find 1-6 backtracking sites. In our simulations that include backtracking, we therefore varied the number of backtracking sites within this range. These parameter values obtained from single-molecule experiments are consistent with the *in vivo* elongation measurements of a *lacZ* template at weak and strong induction of the promoter (19). These experiments show that at low induction RNAPs need 115 sec to transcribe a 2700 nt *lacZ* template, while at strong induction, where backtracking is suppressed due to the dense RNAP traffic, it takes only 40 sec to transcribe this template. If we assume that backtracking is completely suppressed under the latter conditions, backtracking prolongs the total pausing time by 75 sec. If there is a backtracking site every 1000-3000 nt as seen in the single-molecule experiments, this would correspond to 1-3 backtracking pauses and pause durations between 75 sec (if every RNAP makes one backtracking pauses) and 25 sec (for three pauses). These numbers are comparable to what is observed under single-molecule conditions.

At each backtracking site, only a fraction of all RNAPs exhibits a long pause. In single-molecule experiments, this fraction varies between 30 and 80 percent (14, 15). In our simulations we took this fraction to be 50 percent, which we implemented by setting the rate for the transition to the paused state, i.e. the pause frequency $f$ equal to the forward stepping rate $\varepsilon$ at these sites.

The parameters for the dynamics of backtracked RNAPs are the least well-known parameters of our model. Previous estimates based on microscopic models for the dynamics of RNAP led to widely varying results (39, 40). Here we estimate these parameters from the trajectories of RNAPs obtained in single-molecule experiments with single-base resolution. These experiments indicate that single-base steps of a backtracked RNAP occur at a rate about 10x smaller than the elongation step rate (41), from which we estimated the parameter $k_D$ to be ~10 sec$^{-1}$. In addition, we assumed that the transition from the paused non-backtracked state to the backtracked state is given by the first backward translocation step and occurs with the same rate, i.e. $k_b=k_D$. Finally the rate for transcript cleavage was chosen in such a way that RNAPs backtrack over $d\approx$5-20 nt by setting $k_c= k_D/d^2$, since backtracking over distances in this range has been observed both in bulk and single-molecule experiments (19, 38, 41, 42). All simulations that included backtracking were performed for different combinations of the backtracking parameters. While the qualitative results are insensitive to the choice of these parameters, the quantitative details do depend on those parameter values.

## C. Effect of backtracking

To investigate the effect of prolonged pauses with backtracking, we developed an extended version of our model, which is shown in Fig. S3A. During a backtracking pause, a RNAP translocates backwards on both the DNA and the RNA and the newly synthesized 3' end of the transcript is displaced from the active site of the RNAP and extruded (42-45). [There are also other types of prolonged pauses such as RNA hairpin- or protein-mediated pauses, but backtracking appears to be the most common mechanism for prolonging pauses (46).] Backtracking is believed to occur via an intermediate state of the RNAP, which is identical to the RNAP state during short pauses without backtracking and which may be shared with other types of prolonged pauses (43, 46). In our extended model, the entry into the backtracked state is mimicked by a single nucleotide backward move of a paused RNAP, which occurs with rate $k_b$. Backtracking is strongly sequence-dependent (12, 15), which is implemented in our model by allowing this transition only at specific backtracking pause sites. In our simulations we have varied the number of such sites in the range of 1-6 to obtain a frequency of backtracking events similar to the one found in single-molecule experiments. In the backtracked state, the RNAP can make further backward or forward steps (41, 47), provided again that the destination site is not already occupied by another RNAP. Backward and forward steps occur with the same rate $k_D$ to account for the fact that states that differ in the backtracked distance are on average energetically equivalent (47). When a backtracked RNAP returns to the site from where it had entered the backtracked state, it returns to the paused state. An alternative path to escape from the backtracked state is by cleaving the extruded transcript (48). We implemented the latter in our model as a direct transition to the active state with rate $k_c$. During this transition the position of the RNAP is unchanged. At a given instant, the length of the RNA transcript in our model is given by the position where the RNAP was last in the active state. We estimated the parameter ranges for the additional rates $k_b$, $k_D$, and $k_c$ from the experimental literature, see the section on parameter estimation and Table 3.

Fig. S3B and C show results for the transcription rate $J$ and the elongation speed $u$ as functions of the initiation attempt rate $\alpha$ obtained from simulations with 4 backtracking sites (filled symbols). The parameters used in this particular simulation were $k_b=k_D=20$ s$^{-1}$ and $k_c=0.05$ s$^{-1}$. Different choices of the parameters or different numbers of backtracking sites led to qualitatively similar results. These results show that the maximal transcription rate $J_{max}$ is only slightly reduced compared to the case without backtracking (open symbols). Likewise, the elongation speed in the elongation-limited regime (for large $\alpha$) is also reduced only by a few percent; see Fig. S3B. In the initiation-limited regime (for small $\alpha$), where pausing does not affect the transcription rate $J$, backtracking also has no effect on $J$. In this regime, the elongation speed $u$ is considerably reduced compared to the case without backtracking. In addition, the case with backtracking initially exhibits an increase of the elongation speed with increasing initiation attempt rate, i.e. with increasing density of the RNAP traffic. This increase becomes more pronounced as the total amount of backtracking is increased, e.g. by increasing the number of backtracking sites. This increase of the elongation speed can be explained by a reduction of prolonged pausing, as in increasingly dense RNAP traffic it becomes more and more likely that a backward move of the pausing RNAP is not possible, because the upstream site is occupied by the trailing RNAP. Indeed our simulations show that the duration of a backtracking pause (measured as the total time between the first arrival of an RNAP at the backtracking site and the elongation step beyond the site) is strongly reduced as the initiation attempt rate is increased as shown in Fig. S3D. Fig. S3D also shows that the duration of a prolonged pause decreases strongly already for small initiation attempt rates and that it is reduced to the duration of a short pause without backtracking at large initiation attempt rates. The decreased amount of backtracking observed in our simulations is in agreement with recent experimental results by Epshtein and Nudler (19, 49), which show that, both *in vitro* and *in vivo*, backtracking is suppressed in multiple round transcription compared to the transcription by single RNAPs. In particular, Epshtein and Nudler show that the elongation speed of a lacZ mRNA *in vivo* is increased when the promoter is fully induced compared to partial induction (19). While our simulation data shown in Fig. S3C exhibit such an increase in elongation speed with increasing initiation attempt rate $\alpha$, they also show that for very large $\alpha$, the elongation speed decreases again as in the absence of backtracking. This observation implies that for large $\alpha$, the speed-up of elongation by the suppression of backtracking does not overcome the slow-down of elongation due to short pauses and traffic jams behind pausing RNAPs.

In these simulations, our assumptions on RNAP-RNAP interactions are rather conservative. In particular, we have assumed that the backtracked RNAP is not actively pushed forward by the trailing RNAP. This means that the trailing RNAP does not move the leading (backtracked RNAP), but rather passively prevents backward steps of the leading RNAP, when the two are in immediate proximity. Backtracking pauses could be suppressed even more than predicted by our model, if the backtracked RNAP is actively pushed forward by the trailing RNAP. Single-molecule experiments have shown that backtracking is drastically reduced by forward pulling on the RNAP with a force of 8 pN (38). This force is sufficiently below the RNAP stall force, which suggests that a trailing RNAP could exert enough force to push the backtracked RNAP forward actively.

In contrast, short pauses without backtracking (ubiquitous or elemental pauses) are unaffected by force as shown in single-molecule experiments by pulling on either the DNA template or on the RNA transcript with optical tweezers (13, 50). This implies that while active pushing of the trailing RNAP would lead to a stronger reduction of backtracking and to a larger increase of the elongation speed at intermediate initiation attempt rates, it would have little effect on transcription in saturated traffic where short pauses without backtracking dominate.

### D. Constraints on the stepping rate ε

As discussed in section A above, the elongation speeds measured in vivo (17, 20-22) provide a lower bound for the stepping rate ε, which has to be higher than 90-100 $\text{sec}^{-1}$. We have also obtained an upper bound for ε from the requirement of the high transcription rates of rRNA at fast growth. To investigate the effect of the value of the stepping rate ε, we varied ε and, for each value of ε, repeated our analysis of the rRNA elongation speed with and without AT as measured by Vogel and Jensen (18). For each value of $\varepsilon$, we determined the predicted value of the pause duration τ with and without AT and the corresponding maximal value of the transcription rate

$J_{max}$ as described in the main text. The results for the pause duration $\tau$ are plotted in Fig. S6A and show a dependence of the fold-reduction in pause duration on the trial value of the elongation attempt rate $\varepsilon$. In Fig. S6B, we plotted the expected value of the transcription rate, $J_{max}$, for the two sets of pause durations obtained in Fig. S6A. We see that there is about a 2-fold increase in $J_{max}$ due to AT throughout the range of $\varepsilon$ examined. With this increase, the transcription rates (filled circles) are boosted to the range needed to sustain the fastest growth [indicated by the dashed line for 20 min doubling time according to Ref. (51)]. These results show that for $\varepsilon > 110$ sec$^{-1}$, even the presence of AT would not be sufficient to sustain the fastest growth. This reflects the increased effect of the pauses on traffic congestion for faster RNAPs, and allows us to place an upper bound on $\varepsilon$ at ~ 110 sec$^{-1}$. This bound is however dependent on the value of the elongation speed measured in presence of AT. As the values measured by different labs using different strains and methods scatter in the range of 79-90 nt/sec (17, 18, 20, 21), we repeated the analysis using the largest value of 90 nt/sec as measured by Condon et al. (21). In that case, we obtain very similar results, but the reduction of the pause duration due to AT is predicted to be slightly stronger and for the same value of $\varepsilon$ we predict slightly higher values of the maximal transcription rate (between 93 and 71 min$^{-1}$ for $\varepsilon$ between ~100 and 150 sec$^{-1}$, data not shown). Furthermore, using the higher elongation speed of Condon et al. (21) shifts the upper bound on $\varepsilon$ to ~150 sec$^{-1}$.

**E. Location of Rho-dependent termination sites within the *rrn* operons**

The location of Rho-dependent termination sites within the *rrn* operons is not known, but studies of various mutants defective in AT (4, 52-54) indicate that Rho-dependent termination occurs predominantly at two well-defined locations immediately downstream of the boxA loading sequences in the leader region of the operon as well as in the spacer region between the 16S and 23S gene, as indicated in Fig. 1A. These studies provide evidence for termination sites downstream of both boxA sequences: *rrn* operons in a NusB mutant and *rrn* operons with point mutations in both (leader and spacer) boxA sequences exhibit reduced transcription of the 16S gene compared to the leader region and reduced transcription of the 23S gene compared to the 16S gene (52, 54). Based on these observation, these termination sites may be located either within the 16S and 23S genes or upstream of the genes, i.e. in the leader or spacer region of the transcript, respectively. However, electron micrographs of the spatial distribution of RNAPs on the *rrn* operons in a NusB mutant show strong polarity between the 16S and the 23S gene, but little or no termination within these genes (4), suggesting that no strong terminators are located within the 16S and 23S genes. This observation agrees with the earlier observation that in an *rrn* operon with a point mutation in the leader boxA, but a wild-type spacer boxA, the transcription of the 23S gene is not reduced compared to that of the 16S gene (while transcription of the 16S gene is reduced compared to transcription of the leader region of the transcript), indicating that termination occurs only upstream of the 16S probe and not within the 16S gene (53). Likewise, for operons with spacer boxA mutations, Pfeiffer and Hartmann found only a small difference between the expression of an RNA probe in the middle of the 23S gene and expression of full length 23S rRNA, also indicating that there is little termination within the 23S gene (54). Taken together these observations thus suggest that termination occurs predominantly immediately downstream of the box A sequences where the AT complex is loaded, both in the leader and spacer region of the transcript. These locations are indicated by T in Fig. 1A.

**F. Transcription of rRNA in strains with *rrn* deletions**

Even higher transcription rates than those in fast growing wild type cells are necessary in strains with reduced number of *rrn* operons (21). For a strain with four *rrn* operons inactivated, the transcription rate per operon has been estimated to be ~107 min$^{-1}$ (21, 55). At the same time, this strain was found to exhibit an increased rRNA elongation speed of ~135 nt/s (21). These results are in qualitative agreement with our picture that transcription of rRNA at fast growth is limited by transcript elongation. It is however not known by which mechanism this very high elongation rate is achieved and it cannot be easily accounted for in our model given the constraints of the results of all the other *in vivo* and single molecule experiments used to build our model. It has been proposed that the increased speed is an effect of the high transcription rate with densely packed RNAPs pushing each other (55), but based on the dynamics of individual RNAPs revealed by single-molecule experiments (13), we consider this rather unlikely. It is also possible that this strain expresses an unknown factor that affects the RNAP speed or that the RNAP stepping rate is changed by a modified chromosome structure due to the drastic reduction in the number of *rrn* operons. Yet another possibility is that AT is not 100 percent efficient in the wild type (e.g. because one of the Nus factors is limiting) and becomes more efficient in the strain with reduced number of *rrn* operons, as the limiting Nus factor is released from the inactivated operons. Under the last scenario, a certain degree of termination for *rrn* transcription would be expected even for WT cells; this has not been observed, although the existing data may not exclude a small degree of termination.

LEGENDS FOR SUPPORTING FIGURES

**Figure S1: Transcription in dense RNAP traffic.** Transcription rate $J$ (A), RNAP density $\rho$ (B), and elongation velocity $u$ (C) as functions of the initiation attempt rate. The same data are shown as in Fig. 2 (thin line: naïve estimate, solid line: without pauses, dashed line: including pauses), but plotted over a wider range of initiation attempt rates. These plots show that there is an abrupt transition at a critical value $\alpha_c$ (marked by arrowheads in A) between the initiation-limited and the elongation-limited regime. Note that the difference in elongation speed (indicated by arrows in C) between the cases with and without pauses is larger in dense than in dilute RNAP traffic.

**Figure S2: Histograms of RNAP-RNAP distances in the presence and absence of pausing.** (A) and (B) show histograms obtained for dense (saturated) and dilute traffic ($\alpha=50$ min$^{-1}$) for the case without pauses, respectively. (C) and (D) show the corresponding distributions in the presence of pauses with parameters as given by Table 1.

**Figure S3: Effects of backtracking in RNAP traffic.** (A) Extended model including backtracking of paused RNAPs. Active transcription and pausing are modeled in the same way as in the model without backtracking (Fig. 1B), but in addition, at specific backtracking sites, paused RNAPs (white) may enter the backtracked state (light grey), in which the newly synthesized 3' end of the transcript is extruded, with rate $k_b$. Backtracked RNAPs make random backward and forward steps with the same rate $k_D$. When a backtracked RNAP returns to the site from where backtracking initiated, it returns to the paused state. Alternatively, backtracked RNAPs can also return to the active state (dark grey) by cleavage of the extruded transcript, characterized by the cleavage rate $k_c$. (B) and (C): Transcription rate $J$ and elongation speed $u$ as functions of the initiation attempt rate $\alpha$ for a systems with 4 backtracking sites (filled symbols). Open symbols show the corresponding data in the absence of backtracking. (D) shows the reduction of the overall duration of a backtracking pause as the initiation attempt rate $\alpha$ is increased.

**Figure S4: Pausing at random positions versus specific pause sites**. The transcription rate $J$ (A) and the elongation velocity $u$ (C) are shown as functions of the initiation attempt rate $\alpha$ for pausing at any spatial position (open symbols, solid line) and for pausing at specific pause sites (black dots, dashed lines, two different templates). In the homogeneous case, pauses (without backtracking) occurred stochastically at any position with rate $f=0.1$ sec$^{-1}$. For case with pause sites, a number of pause sites were chosen randomly, on average one per 100 nt as suggested by single molecule experiments (see Supporting Text), at these sites RNAPs paused with rate $f=10$ sec$^{-1}$, so that spatial average of the pause frequency is the same in both cases. B and D show histograms of the maximal transcription rate $J_{max}$ and the corresponding elongation speed $u_{min}$ obtained from 49 randomly generated templates with pause sites. The values of these quantities scatter around those for the case with homogeneous pausing (indicated by triangles).

**Figure S5: RNAP "traffic jams" induced by pausing.** (A) Fractions of RNAPs that are paused (circles) or jammed behind a paused RNAP (squares) and (B) average number of RNAPs jammed behind a paused RNAP as functions of the initiation attempt rate $\alpha$. Jammed RNAPs are those that cannot step forward because either the forward site is occupied by a paused RNAP or by another RNAP jammed behind a paused RNAP. While the fraction of paused RNAPs is constant as a function of $\alpha$, the fraction of jammed RNAP strongly increases.

**Figure S6: Effect of antitermination.** The plot shows the pause duration $\tau$ (A) and the maximal transcription rates $J_{max}$ (B) in absence and presence of antitermination as predicted by matching the pause duration to the measured elongation speeds (18) as a function of the stepping rate $\varepsilon$. The dashed lines indicate the transcription rates of rRNA per operon necessary for fast growth with doubling times of 30 and 20 minutes, respectively (51).

**Table S1: RNAP densities and transcription rates for simulated mutants defective in AT compared to experimental data for NusA and NusB mutants from ref. (4).**

|  | Number of RNAPs per gene (mean ± std deviation) | | Transcription rate (23S gene) | Doubling time | Fraction of RNAP-RNAP distances >250 nt [%] (**) | |
|---|---|---|---|---|---|---|
|  | 16S | 23S |  |  | 16S | 23S |
| WT (exp.) | 20.5 ± 5 | 31.0 ± 10 | - | 24 min | 4.4 | 4.7 |
| Full AT (simul.) | 20.5 ± 2.2 | 35.0 ± 3.0 | 58 min$^{-1}$ | ~24 min(*) | 1.7 | 2.3 |
| NusB mutant (exp.) | 16.4± 3 | 16.2 ± 4 | - | 54 min | 8.5 | 21.6 |
| NusA mutant (exp.) | 14.6 ± 3 | 16.9 ± 3 | - | 38 min | - | - |
| 40% AT (simul.) | 16.1 ± 2.8 | 17.6 ± 3.8 | 26 min$^{-1}$ | ~36-38 min(*) | 9.3 | 26.2 |

(*) Theoretical doubling times were estimated from the transcription rate obtained in our simulations using the relation of transcription rate per *rrn* operon and growth rate (51, 56).

(**) Obtained from the histograms in ref. (4) and in Fig. 3B and C.

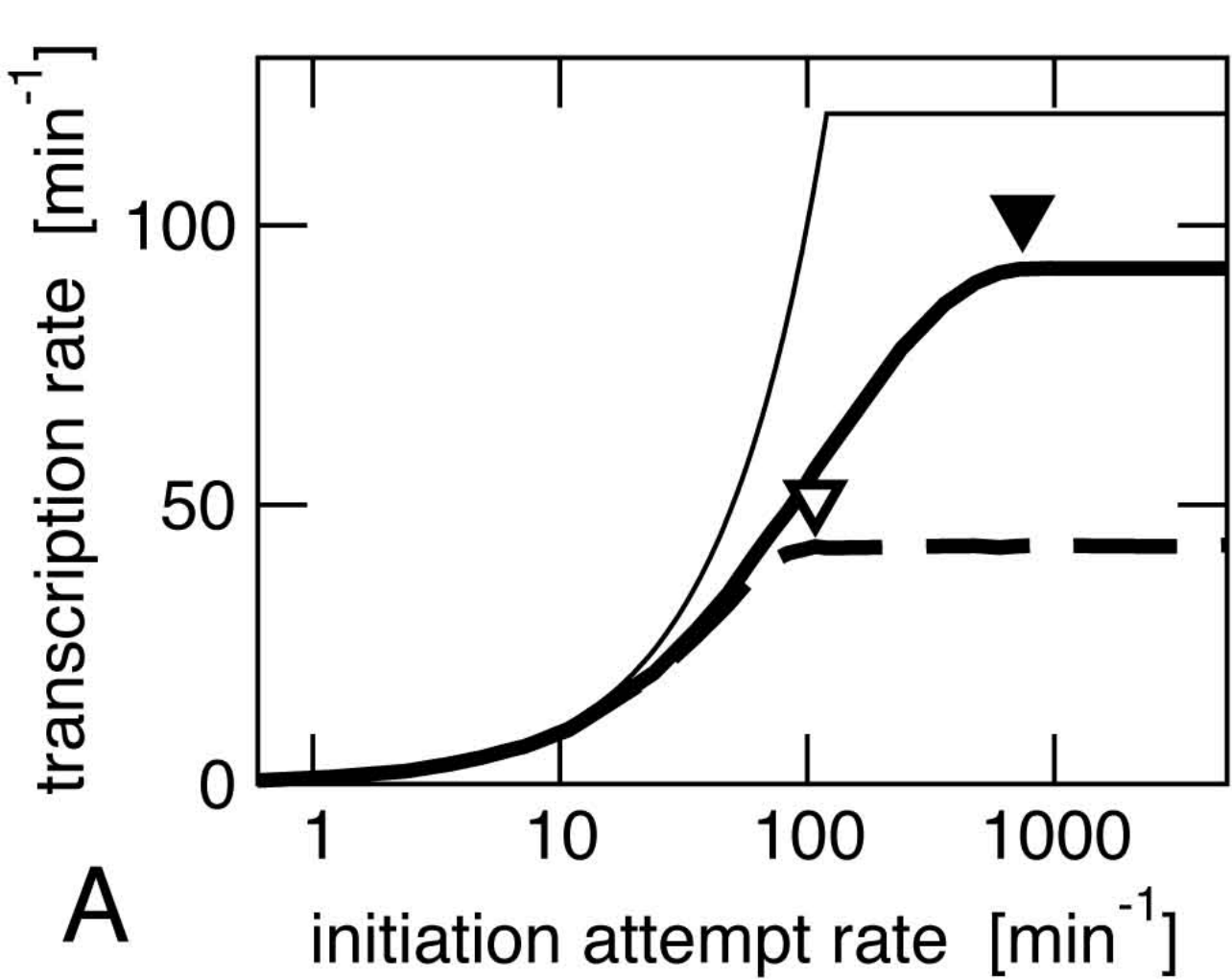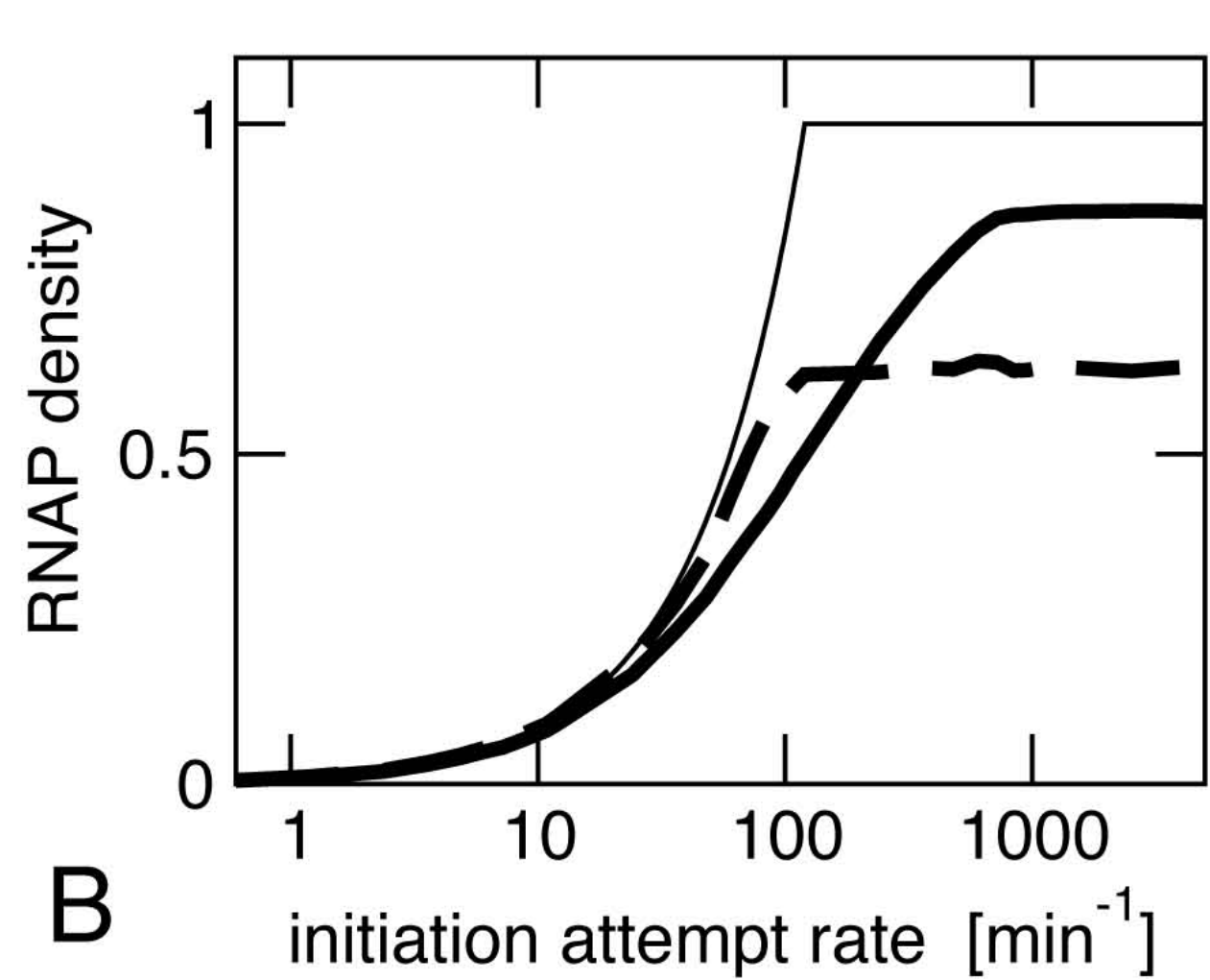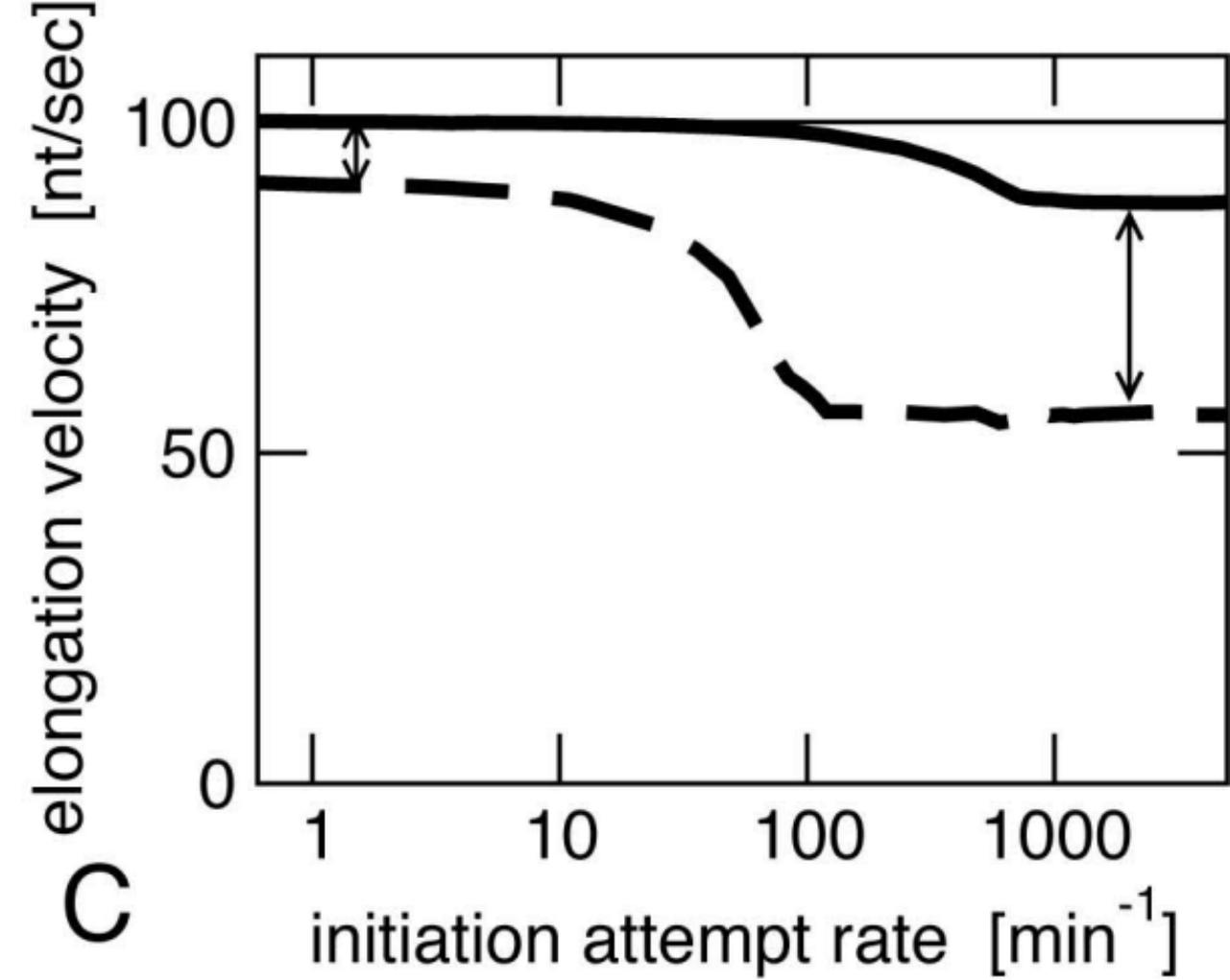

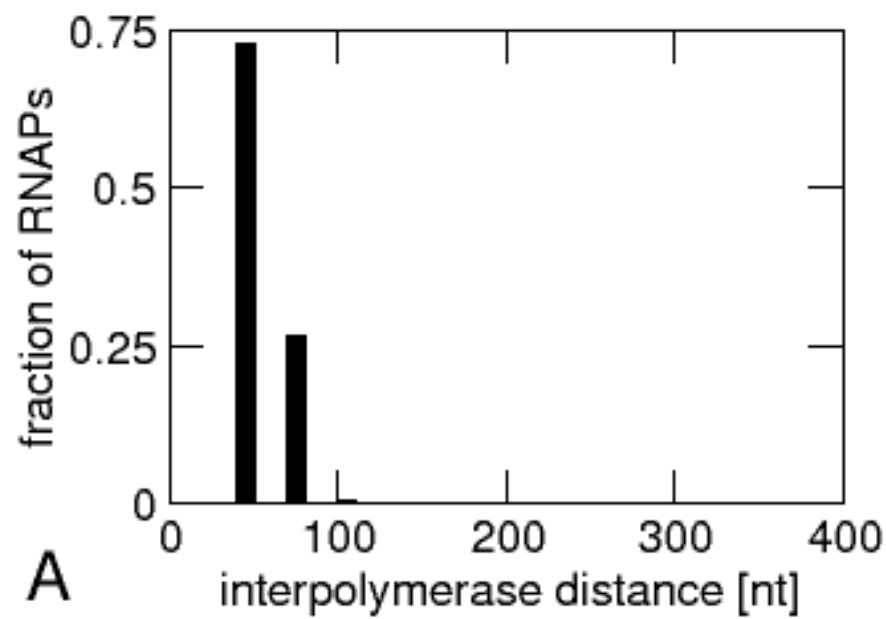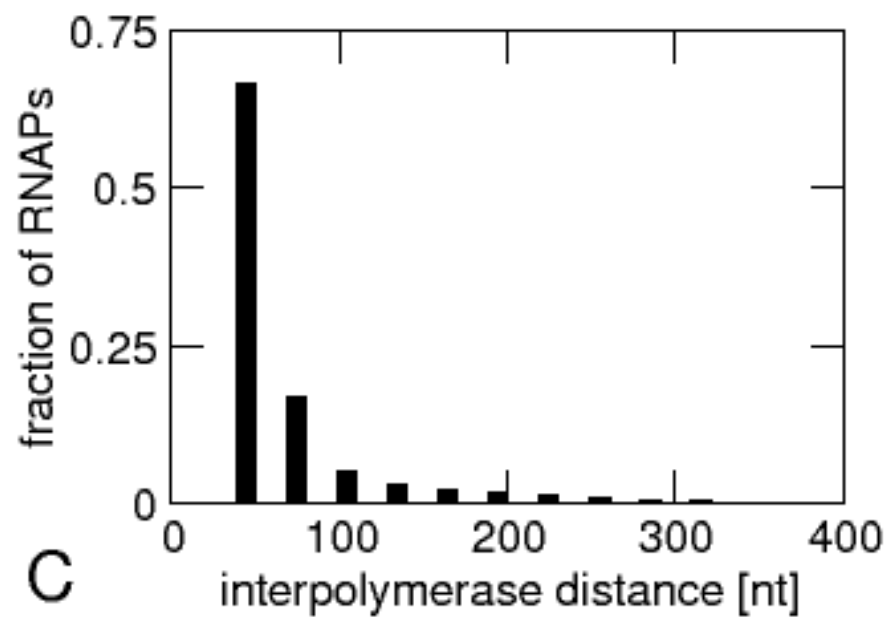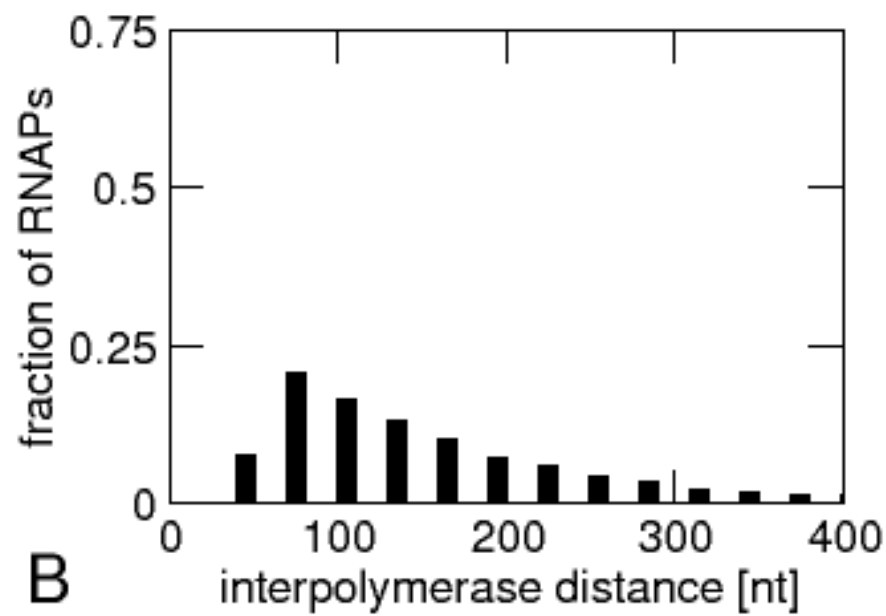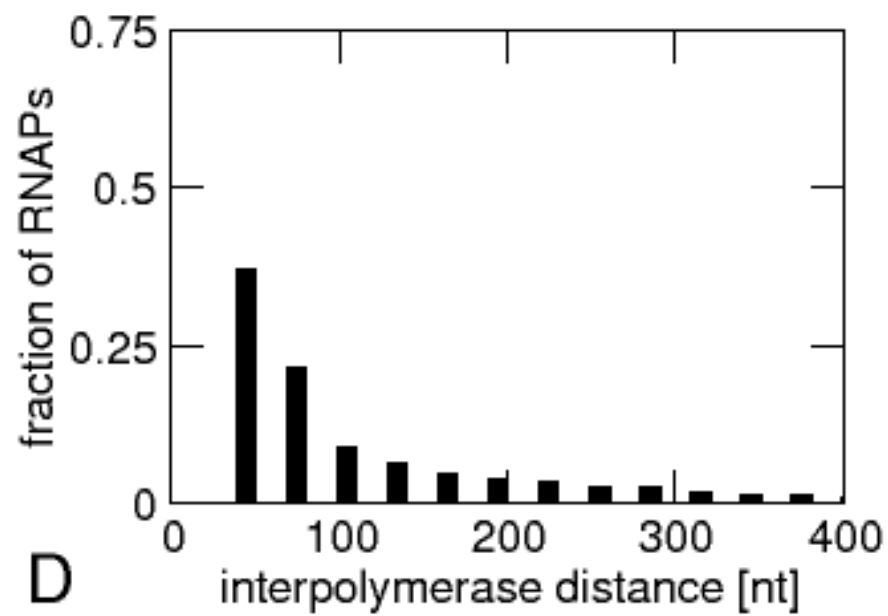

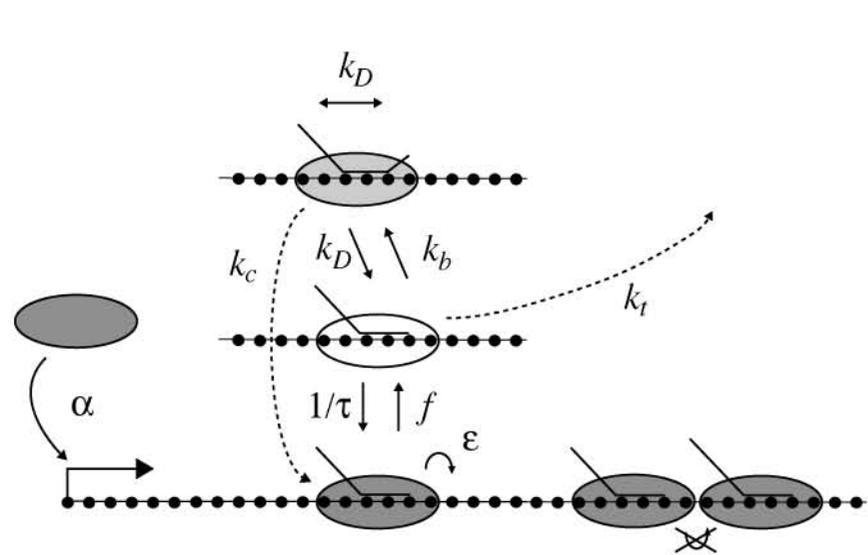
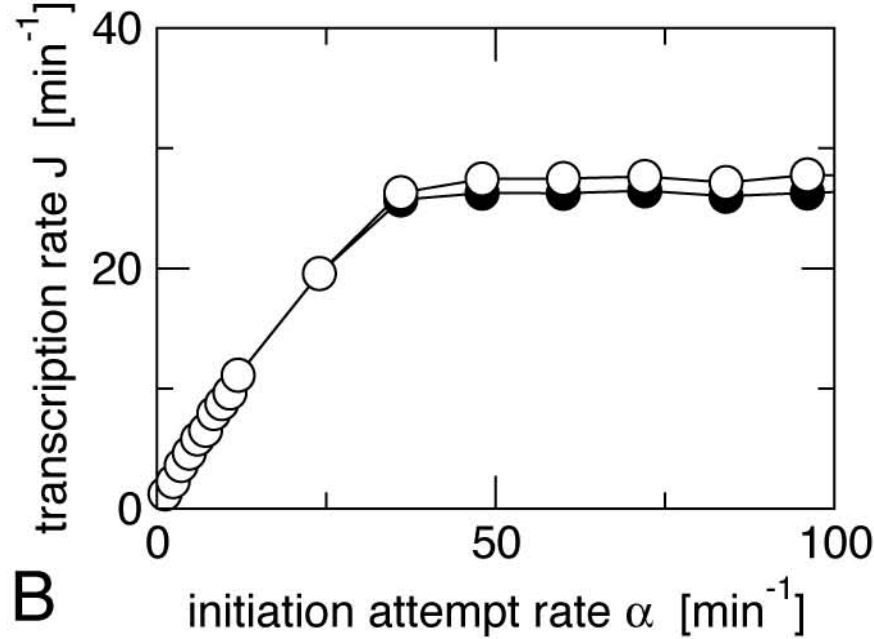
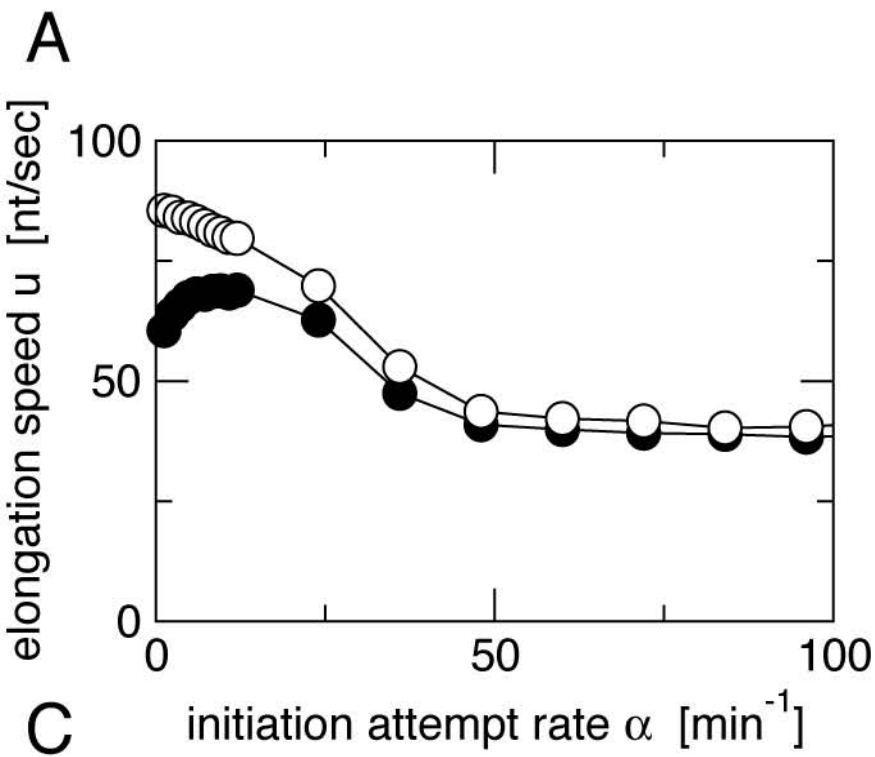
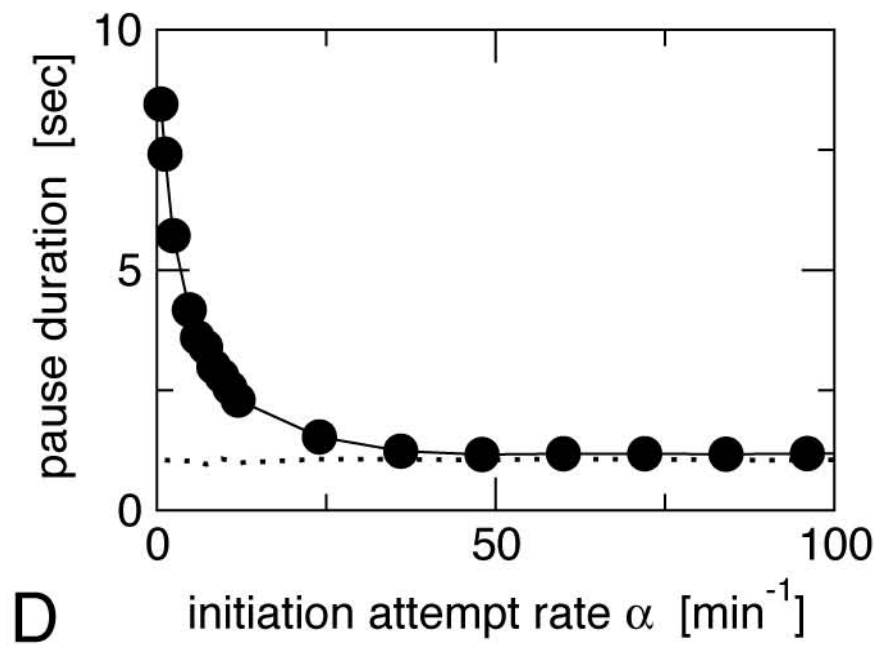

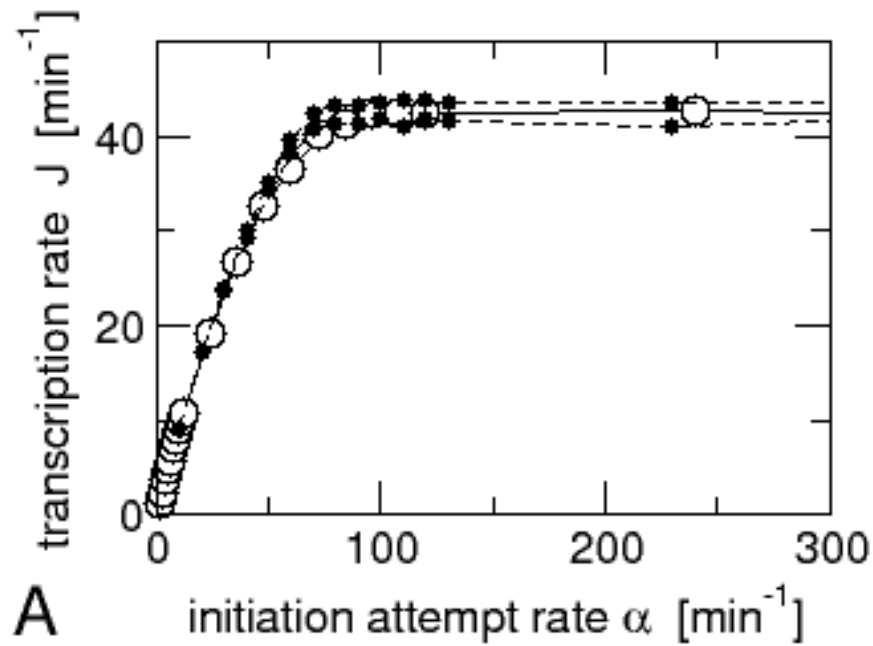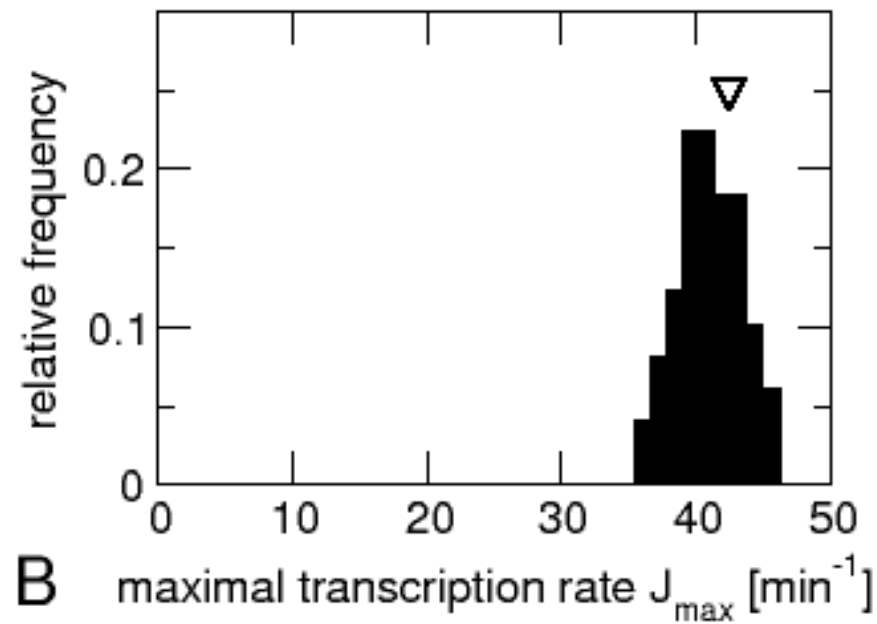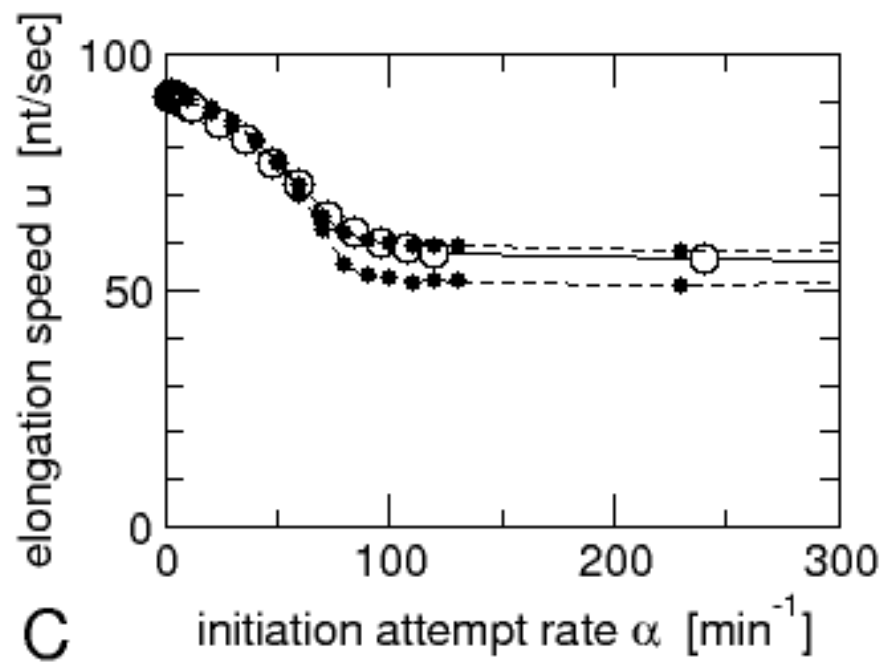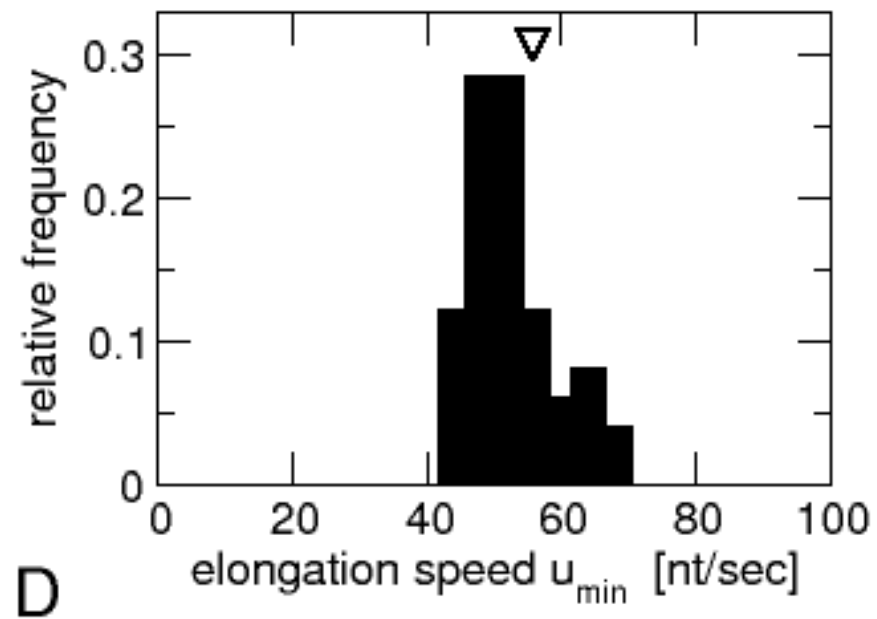

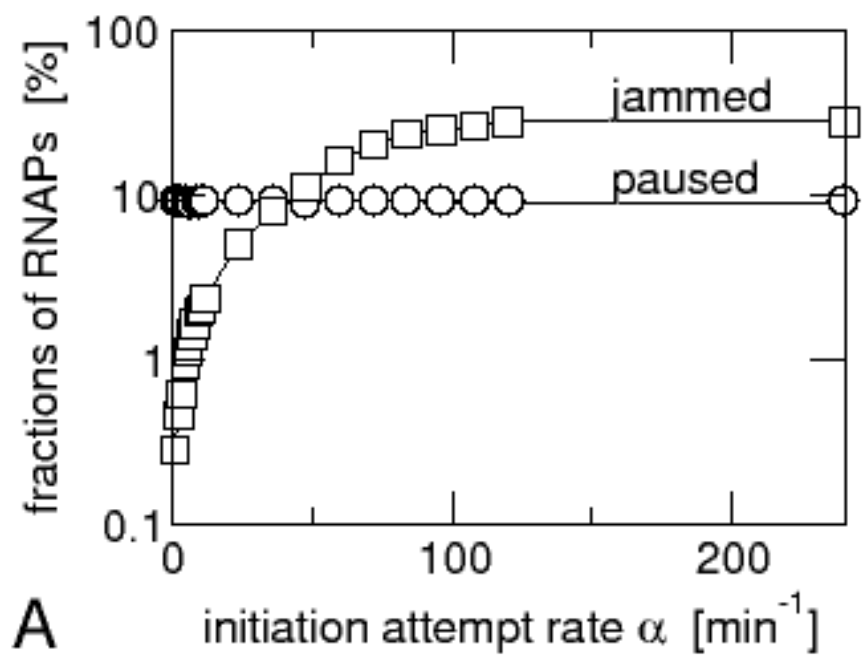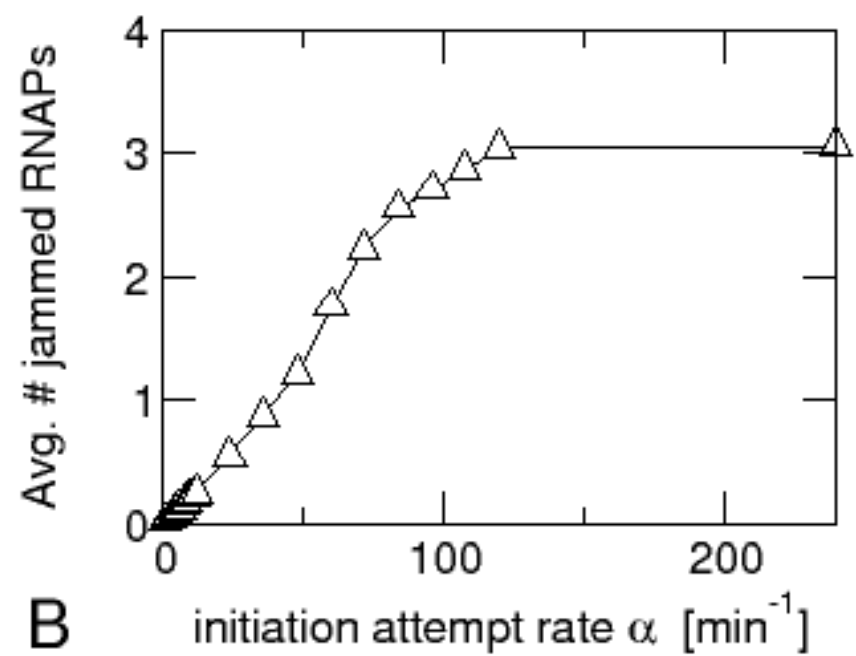

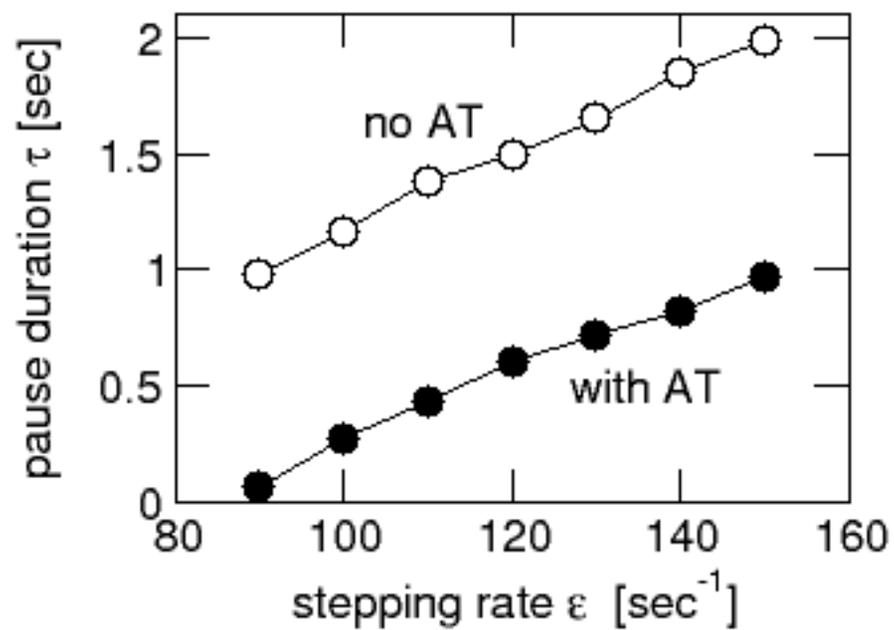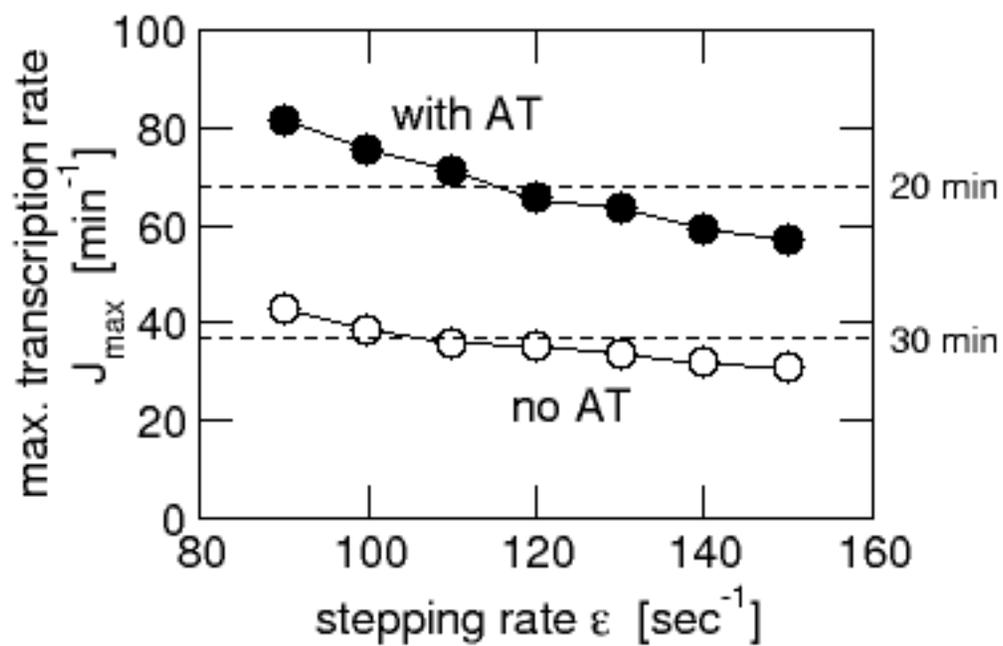